\documentclass{an}

\usepackage{graphicx}
\usepackage{times}
\newcommand{\p}{\phantom{1}}
\newcommand{\m}{$-$}

\begin{document}

\Pagespan{789}{}
\Yearpublication{2014}%
\Yearsubmission{2014}%
\Month{08}%
\Volume{999}%
\Issue{88}%

\title{Astronomical orientation analysis of three proto-historical
  sites in Friuli - Italy}

\author{F. Patat\inst{1}\fnmsep\thanks{Corresponding author:
  \email{fpatat@eso.org}\newline}
\and S. Corazza\inst{2}
}

\titlerunning{Astronomical orientation analysis of three proto-historical sites}

\authorrunning{F. Patat \& S. Corazza}

\institute{
European Southern Observatory - 
     K.-Schwarzschild-Str. 2, 85748 Garching b. M\"unchen, Germany
\and
Laboratorio di Archeologia - Universit\`a di Udine - v. Florio 2, 33100 Udine, Italy
}

\received{15 August 2014}
\accepted{later}
\publonline{later}


\abstract{In this paper we present the results of an
  archaeoastronomical survey of three proto-historical sites located
  in the high Friulian plain (Galleriano, Gradisca and Savalons),
  dating from the end of the Early Bronze Age (1900 B.C.) to the end
  of the Late Bronze Age (950 B.C.)  These structures, commonly
  indicated as castellieri, are earthworks of quadrangular shape, with
  sides ranging from 140 to 250 m. At present the perimetrical earthen
  embankments reach a maximum base width of 18 m and an elevation of
  more than 5 m the surrounding plain in their best preserved parts.
  These three sites were often reported in the literature to have the
  corners aligned to the cardinal directions. Aveni and Romano
  (\cite{ar86}) included two of them (Galleriano and Gradisca) in
  their study of earthworks in Veneto and Friuli (Italy), tentatively
  proposing astronomically relevant alignments for some sides and
  diagonals. Inspired by this pioneering work and by the renovated
  archaeological interest for these sites, we obtained digital
  elevation models of the earthworks and re-analyzed their
  orientation. Our study does not confirm the presence of systematic
  and statistically significant solar or lunar alignments for these
  sites.}

\maketitle

\section{\label{sec:intro} Introduction}

The region of Friuli (Italy) is rich of proto-historical earthworks
dating from the end of the Early Bronze Age (1900 B.C.) to the end of
the Late Bronze Age (950 B.C.). The reader is referred to C\`assola
Guida (\cite{cg80}, \cite{cg06}) and Simeoni \& Corazza
(\cite{simeoni}) for thorough reviews on the subject. The sites are
distributed both in the high Friulian plain and in the pre-Alps, to
the North and North-East of the region. Those located in the plain are
particularly interesting for the archaeoastronomical analysis, because
they could be planned and built without being subjected to strong
geomorphological constraints, as opposed to the cases when the sites
are erected on natural terrain elevations. 

Although many of these earthworks were found in the upper Friulian
plain (Quarina \cite{quarina}), only three of them are sufficiently
well preserved to allow an astronomical orientation analysis. In all
other cases modern activities modified their original morphology so
severely that their only remnants are mostly
buried below the surface (C\`assola Guida \& Corazza \cite{cgc05}) and
no orientation analysis is possible at the moment.

\begin{table*}
\tabcolsep 2.8mm
\caption{\label{tab:data} Geographical and geometrical parameters of the sites}
\centerline{
\begin{tabular}{cccccccccc}
\hline
 	& \multicolumn{4}{c}{Geodetic position} & Elevation & Sides & Perimeter & Area & Volume\\
         \cline{2-5} \\
Site   	& Longitude & Latitude & Easting & Northing & (m)  & (m) & (m) & (m$^2$)  & (m$^3$)\\
       	& (deg)          & (deg)      & (m)     & (m)  &              &        &        &      &\\
	\hline
Galleriano & 13.1130E & 45.9824N & 353840 & 5093818 & 53    & 245$\times$160 & 830 & 42900 & $>$6000\\       
Gradisca   &  12.9692E & 46.0069N & 342772 & 5096821 & 65    & 170$\times$140 & 620 & 26100 & $>$12300\\
Savalons  &   13.0629E & 46.0607N & 350169 & 5102618 & 104 & 180$\times$180 & 700 & 37900 & $>$9900\\
\hline
\multicolumn{10}{l}{\footnotesize Notes: Coordinates refer to the approximate geometrical centers of the structures and were determined from the survey data.}\\
\multicolumn{10}{l}{\footnotesize Longitude and Latitudes are in the WGS84. Easting and Northing are in the UTM grid, zone 33N.}\\
\end{tabular}
}
\end{table*}

\begin{figure}
\centering
\includegraphics[width=8cm]{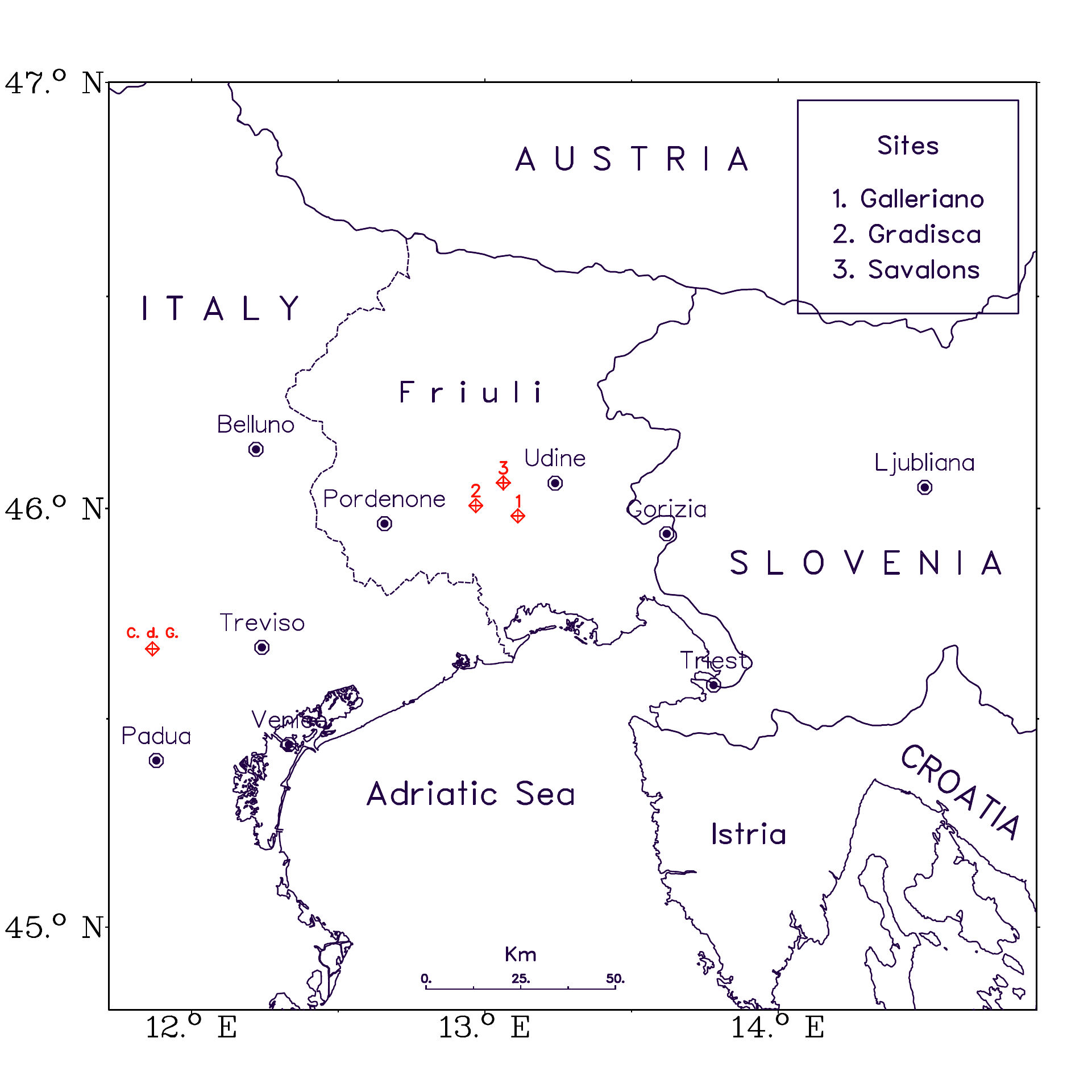}  
\caption{\label{fig:map} Area map showing the locations of the three
  proto-historical sites: Galleriano di Lestizza (1), Gradisca di
  Sedegliano (2) and Savalons di Mereto (3).}
\end{figure}

The three sites discussed in this paper are located in the
surroundings of the villages of Galleriano di Lestizza, Gradisca di
Sedegliano and Savalons di Mereto, in the province of Udine (see
Fig.~\ref{fig:map}) and have been object of several archaeological
campaigns (Corazza \cite{susi00}; C\`assola Guida \& Corazza
\cite{cgc03b, cgc03c, cgc04a, cgc04b, cgc05, cgc09}). The accurate
positions for the three sites are given in Table~\ref{tab:data}.
These structures are commonly indicated as {\it castellieri} (we will
conform to this nomenclature throughout this paper) and are earthworks
of quadrangular shape, with sides ranging from 140 to 250 m. At
present, the earthen embankments reach a maximum base width of 18 m
and an elevation of more than 5 m above the surrounding plain in their
best preserved parts. Portions of the castellieri were destroyed,
especially by modern agricultural activities. In particular, the
perimetrical embankment was broken in one or more points to give easy
access to cultivations. The recent archaeological digs are starting to
clarify the role of these sites, proving that they were
inhabited. Many questions remain unanswered (C\`assola Guida
\cite{cg06} and references therein) but a recent finding in the
castelliere of Gradisca allowed the archaeologists to put a firm
estimate on its construction date.  In 2004 several human skeletons
were found buried in the earthen embankment and their radiocarbon
dating led to the conclusion that the foundation of the site took
place at the end of the Early Bronze Age (c.a 1900 B.C.; Borgna \&
C\`assola Guida \cite{borgna}). More uncertain is the dating of the
other two sites, but there are indications that they might be slightly
more recent, judging from the ceramic remnants (C\`assola Guida \&
Corazza \cite{cgc03b, cgc05, cgc09}). However, only future digs
reaching the base of the embankments (buried below the current level
of the surrounding plain) will cast more light on this issue, allowing
for a more accurate dating.

What is clear from the available data is that the construction of the
sites took place in different and subsequent stages during the Bronze
Age, and some modifications were also introduced during the Roman
occupation (second century B.C.).

A characteristic property of these sites is the presence, in their
neighborhoods, of earthen burial mounds (C\`assola Guida \& Corazza
\cite{cgc03a}). In 2008 the University of Udine completed an
excavation campaign at the mound of Mereto di Tomba, located about 2
km SSW of the castelliere of Savalons. This led to the discovery of a
human male skeleton, deeply buried below the mound. The radiocarbon
dating provided an age between 3620 and 3830 years (Borgna, Corazza \&
Simeoni \cite{borgna13}).

While the Histrian castellieri are characterized by a circular shape,
the Friulian ones are quadrangular, hence leading to considerations
about their orientation.  They are rather frequently reported to have
the corners aligned to the cardinal points (Quarina \cite{quarina}; di
Caporiacco \cite{caporiacco}; C\`assola Guida \cite{cg80}; Menis
\cite{menis}; C\`assola Guida \& Corazza \cite{cgc05}; Simeoni \&
Corazza \cite{simeoni}). To the best of our knowledge, the only study
specifically devoted to an orientation analysis of these sites in
connection to astronomically relevant directions is the one by Aveni
\& Romano (\cite{ar86}; hereafter AR86), who undertook a survey of
   {\it motte} and castellieri mostly in the neighboring Veneto
   region.

Motte, also known as {\it mutere}, are earthen structures of roughly
conical form, with heights spanning from 2 to 8 m and diameters
between 6 and 60 m. The study by Aveni and Romano is mostly devoted to
the alignments between mutually-visible motte, but it includes also
two Friulian castellieri, i.e. Galleriano and Gradisca. The connection
to astronomically significant directions is not very clear for these
two sites. For the minor diagonal of the latter, AR86 found a slightly
deviating equinoctial alignment (they report an azimuth of 85.0
degrees, corresponding to a refraction corrected declination of +3.0
degrees). For the remaining directions (sides and diagonals) they
tentatively considered minor/major lunar standstills and stellar
alignments (namely the rising of Aldebaran, for the minor diagonal of
Galleriano); however, they did not attach very much confidence to
these findings and they reported them only for the sake of
completeness.

Stimulated by the pioneering work of AR86, the renovated interest shown
by the archaeological community and the fruitful digs of these last
fifteen years, we decided to investigate in a more thorough way the
two Friulian sites of Gradisca and Galleriano, and to extend the
analysis to the unstudied castelliere of Savalons.

The paper is structured as follows. In Sect.~\ref{sec:data} we
describe the methods used to acquire the data. Sect.~\ref{sec:morph}
presents the general morphology of the sites, which are then analysed
in Sect.~\ref{sec:orient}. The results are finally discussed in
Sect. \ref{sec:disc}, which also recaps our
conclusions. Appendix~\ref{sec:appa} provides a description of the
numerical methods used for horizon altitude calculation, refraction
correction and statistical error estimates. Appendix~\ref{sec:appb}
presents the digital elevation models (DEM) for the three sites.

\section{\label{sec:data}Survey data}

Rather than measuring the azimuths of specific directions at each
site, as it was done by AR86, we preferred to go for a full
three-dimensional survey of the earthen structures. Besides enabling a
more accurate morphological study, this methodology allows one to
determine azimuths using mathematical fitting procedures, as opposed
to the direct measurement of sight lines that necessarily imply a
somewhat arbitrary choice of two reference points defining the
alignment. This choice becomes particularly difficult when one is to
determine the diagonals of a quadrangular structure with sides of
100-200 m and rounded corners, or its symmetry axis. This is made even
more cumbersome by the presence of vegetation, as is the case in
Galleriano and Savalons.  Galleriano and Gradisca were mapped in 2004
and 2005 for the University of Udine by the professional surveyor
G. Meng, using a Wild-Leica Total Station. A total of 687 and 588
points were surveyed for Galleriano and Gradisca, respectively. The
relative horizontal/vertical measurements were translated into an
absolute geodetic grid by means of a number of trigonometric fiduciary
points visible from the two sites. Finally, the reconstruction of the
three-dimensional geometry of the earthworks from the sparsely
distributed data points was performed by linear interpolation
within triangles, determined via Delaunay triangulation. The results
were stored as digital images with a nominal resolution of 0.5 meters
per pixel.

No modern data suitable for our analysis were available for the
castelliere of Savalons. Therefore, we decided to survey the site
taking advantage of the Light Detection And Ranging (hereafter LIDAR)
technology. LIDAR is a remote sensing technique that makes use of a
scanning laser beam to get range information on a distant target, and
it is based on the measurement of the travel time of laser
micro-pulses and on the properties of the light reflected back by the
target. In our specific case, the advantage of a laser scan over a
classical total station survey is the much higher spatial resolution and the
definitely shorter time it takes to cover a 4
hectares site.  The measurements were taken from an helicopter in
November 2007 by the HELICA company, using an Optech laser scanning
system operating at 1,064 nm. The horizontal positions of the sampled
points are derived from the simultaneous use of 12 channels,
dual-frequency GPS receiver on board of the aircraft coupled to ground
GPS beacons located within 30 km. They are then converted to a
geodetic absolute grid during the data post-processing. Typical root
mean square (rms) vertical accuracies are better than 5 cm, while the
rms absolute horizontal accuracy is less than 20 cm. The survey was
completed in a few seconds, from an elevation of about 1,000 m. Even
though higher spatial resolutions are possible, both because of the
nature of the site and the purpose of this work the scan was performed
with a resolution of 30 cm over a rectangular area of 467 m $\times$
398 m, for a total of about 1,723,000 data points.

\begin{figure}
\centering
\includegraphics[width=8cm]{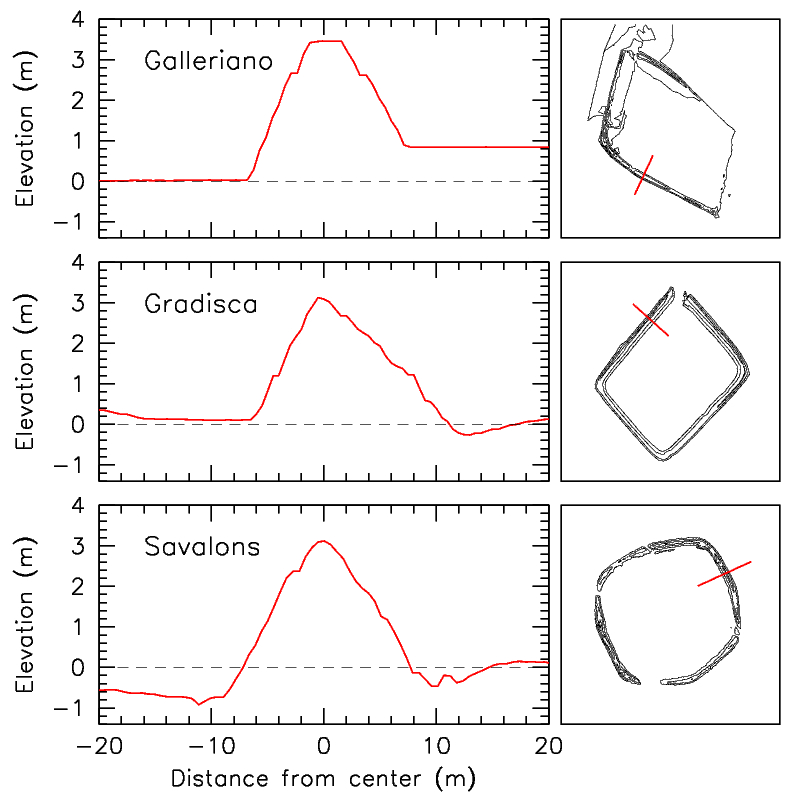}  
\caption{\label{fig:trac} Representative cross sections of the
  embankments for the three sites (from the outside to the inside). 
  The right panels are contour plots extracted from the DEMs 
  presented in this paper (see Sect.~\ref{sec:data}).}
\end{figure}

The only drawback of a airborne LIDAR survey, when compared to ground
based measurements, is the disturbance introduced by vegetation,
possibly preventing the laser from reaching the soil, and hence
giving false altimetry. The problem was partially mitigated by
running the survey when the foliage had fallen. Moreover, based on the
different properties of light reflected by the soil or by vegetation,
during the data processing it was possible to disentangle ground from
non-ground points. This reduced the number of suitable measurements to
about 590,000, corresponding to an average resolution of 56 cm. In the
analysis we used only the ground points. The reconstruction of the
three-dimensional geometry of the site was carried out in a similar
way as for the total station surveys, i.e. via Delaunay triangulation
and linear interpolation. The comparison between the DEM
and aerial ortho-photography shows that this procedure
removes vegetation quite effectively.

The resulting DEMs for the three sites are presented in
Appendix~\ref{sec:appb} (Figs.~\ref{fig:map_gall}, \ref{fig:map_grad}
and \ref{fig:map_sava}).  They are geo-referenced to the World
Geodetic System 1984 (hereafter WGS84), so that direct azimuth
measurements are possible without any additional correction for grid
rotation (horizontal and vertical axes are aligned to the true North
and true East, respectively).

In all cases we validated the whole procedure by directly measuring
the absolute azimuth of some well defined and straight features in
each site using a RK~76 Zeiss theodolite and sun fixes. The agreement
is always within 0.1 degrees and the discrepancies are well explained
by the difficulty of placing the field check points at the
correct positions on the DEMs (see Sect.~\ref{sec:orient}).

\begin{figure}
\centering
\includegraphics[width=8cm]{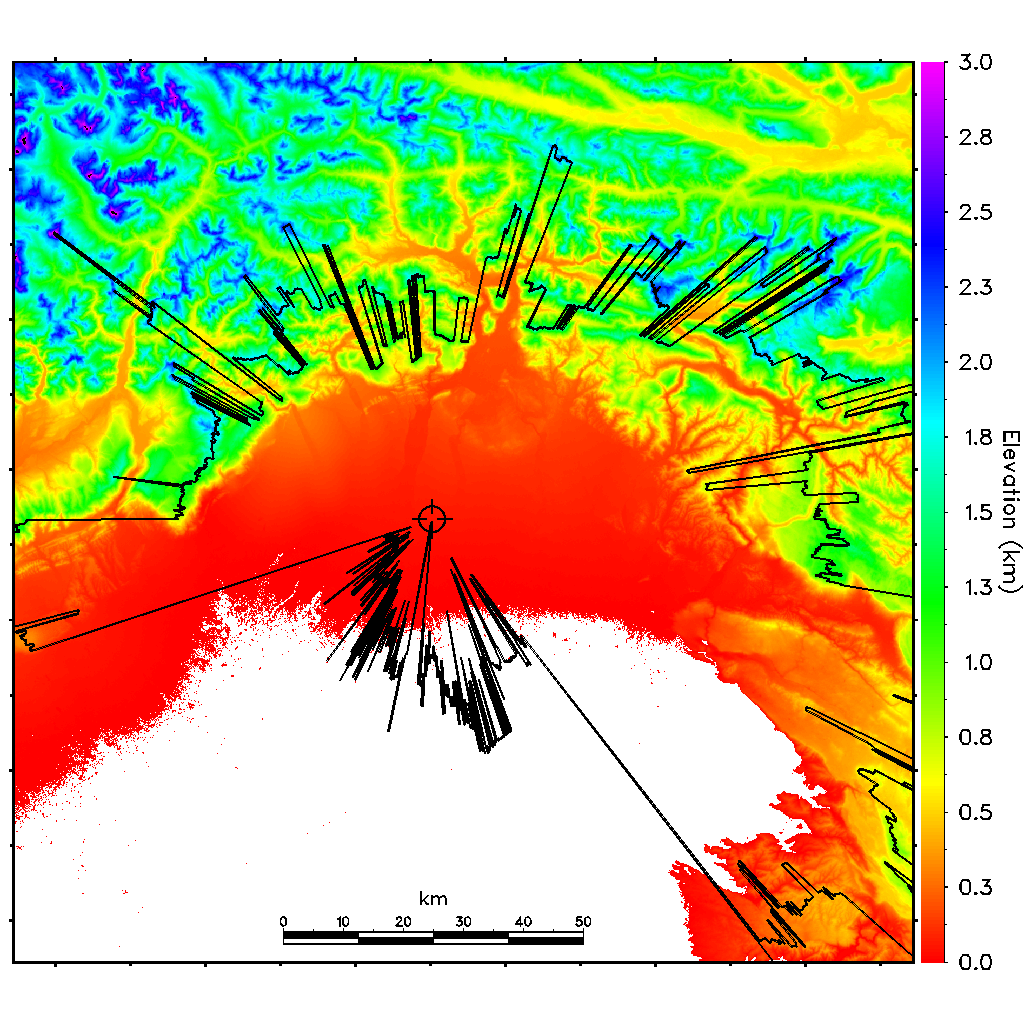}  
\caption{\label{fig:horizon} Example horizon calculation for the site
  of Gradisca. The circle in the center marks the site position.}
\end{figure}

\section{\label{sec:morph} General morphology}

The sides of the quadrangular structures range from 140 to 245 m,
while the enclosed areas vary from 2.6 to 4.3 hectares. Only
lower limits of the original volumes can be estimated; they range
from 6$\times$10$^3$ m$^3$ (Galleriano) to almost 1.3$\times$10$^4$
m$^3$ (Gradisca). Adopting the typical density of an earthen mixture
($\sim$2$\times$10$^3$ kg m$^{-3}$), these numbers imply masses that
range from 1.2$\times$10$^7$ to 2.6$\times$10$^7$ kg. Representative
cross sections of the embankments  are shown in
Fig.~\ref{fig:trac} for the three sites. The full width of these profiles at the current
ground level is 15.8, 17.5 and 15.1 m (full width half maximum is 9.8,
9.3 and 8.9 m), while the peak elevation above the surrounding plain
is 3.4, 3.1 and 3.2 m for Galleriano, Gradisca and Savalons,
respectively. The largest values are reached in Savalons, where the
maximum base width is about 18 m and the elevation slightly exceeds 5
m.

Although the three sites are commonly classified as quadrangular
structures, they show clear morphological differences. The most
regular shape is the one of Gradisca, which is close to a
parallelogram with very rectilinear sides forming angles that deviate
by less then 10 degrees from the right angle
(Fig.~\ref{fig:map_grad}). The shorter sides (NE and SW) are parallel
to within $\sim$0.1 degrees, while the longer sides deviate by $\sim$4
degrees from parallelism (see Sect.~\ref{sec:grad}). The case of
Galleriano is less regular: the sides are only partially rectilinear
and show a pronounced bending at the western and northern corners
(Fig.~\ref{fig:map_gall}). The SE side and the southern portion of the
NE side were almost completely leveled by modern agricultural
activities. Finally, Savalons presents clear bends in all
sides\footnote{Romano (\cite{romano}) proposed a possible practical
  way of designing a site like Savalons, using four sticks and a
  string (his Fig.~64) Nevertheless, this geometrical method produces
  a quadrilateral with curved sides and rounded corners, while the
  data presented here suggest the existence of rectilinear segments,
  especially in the SW, SE and NE sides (see Fig.~\ref{fig:map_sava}
  and next section). We note that Romano (\cite{romano}, p. 223, his
  Fig.~65) proposed a similar approach for the site of Gradisca; however, 
  this is in contradiction with the very rectilinear sides of the
  embankment.} (Fig.~\ref{fig:map_sava}).

\begin{figure}
\centering
\includegraphics[width=8cm]{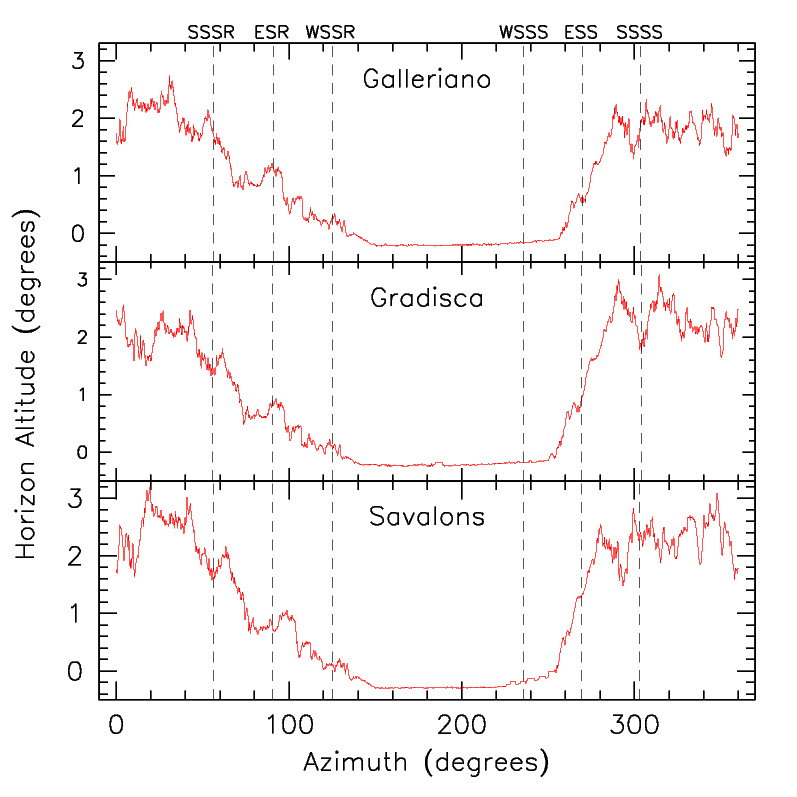}  
\caption{\label{fig:horprof} Synthetic horizon profiles for the three
  sites. The vertical dashed lines marked the solar reference
  positions (see Table~\ref{tab:sun} and Sect.~\ref{sec:orient})}.
\end{figure}

Geographical and geometrical parameters for the three sites are
recapped in Table~\ref{tab:data}. Both sets of parameters were
estimated from the survey data presented in this work. The length of
the sides was measured along a straight line, corner to corner. Given
the rounded shape of Savalons, the side length is only indicative. The
perimeter was estimated as the length of a closed polygonal line along
the maximum elevation of the embankments. The site area was derived as
the plain surface enclosed by the polygonal perimeter. The volume was
computed by integrating the elevation data of the embankments above a
certain threshold altitude; therefore, the reported values are only
lower limits. For Galleriano the estimate is particularly uncertain,
because of the bad state of the NE and SE portions of the earthwork.

\begin{table*}
\caption{\label{tab:sun} Refraction and horizon altitude corrected
  azimuths for sunrise (SR) and sunset (SS) for 1500 B.C.. The
  uncertainties are one standard deviation (SD). All values are in decimal
degrees.}  
\tabcolsep 0.90mm
\centerline{
\begin{tabular}{ccccccccccccc}
\multicolumn{13}{c}{Sun center}\\
\hline
Site & WSSR & h & WSSS & h & ESR & h & ESS & h & SSSR & h & SSSS & h \\
\hline
GAL &  125.18$\pm$0.24 & +0.26  & 235.49$\pm$0.27 &\m0.17 & 90.71$\pm$0.17 & +1.14  & 269.97$\pm$0.19 & +0.57 & 56.09$\pm$0.18 & +1.75  & 303.65$\pm$0.17 & +1.94\\
GRA &  124.93$\pm$0.25 & +0.09  & 235.48$\pm$0.27 &\m0.17 & 90.41$\pm$0.17 & +0.88  & 269.55$\pm$0.17 & +0.92 & 55.59$\pm$0.19 & +1.39  & 303.75$\pm$0.18 & +1.88\\
SAV &  124.97$\pm$0.25 & +0.09  & 235.45$\pm$0.27 &\m0.18 & 90.24$\pm$0.18 & +0.74  & 269.12$\pm$0.16 & +1.28 & 56.05$\pm$0.18 & +1.76  & 303.12$\pm$0.16 & +2.38\\
\hline
CDG &  124.34$\pm$0.27 &\m0.14  & 235.07$\pm$0.24 & +0.24 & 89.72$\pm$0.20 & +0.32  & 268.90$\pm$0.15 & +1.48 & 54.47$\pm$0.23 & +0.45  & 302.81$\pm$0.16 & +2.43\\
\hline 
\multicolumn{13}{c}{}\\
\multicolumn{13}{c}{Sun lower limb}\\
\hline
Site & WSSR & h & WSSS & h & ESR & h & ESS & h & SSSR & h & SSSS & h \\
\hline
GAL &  125.63          & +0.26  & 235.02          & -0.16 & 91.07          & +1.10  & 269.52          & +0.61 & 56.46          & +1.69  & 303.32          & +1.85\\
GRA &  125.40          & +0.09  & 235.02          & -0.17 & 90.77          & +0.84  & 269.18          & +0.89 & 56.02          & +1.38  & 303.31          & +1.87\\
SAV &  125.45          & +0.10  & 235.00          & -0.18 & 90.62          & +0.72  & 268.68          & +1.30 & 56.47          & +1.73  & 302.77          & +2.30\\
\hline
CDG &  124.79          & -0.14  & 234.70          & +0.18 & 90.12          & +0.32  & 268.53          & +1.46 & 54.85          & +0.41  & 302.38          & +2.42\\
\hline
\multicolumn{13}{c}{}\\
\multicolumn{13}{c}{Sun upper limb}\\
\hline
Site & WSSR & h & WSSS & h & ESR & h & ESS & h & SSSR & h & SSSS & h \\
\hline
GAL &  124.70          & +0.24  & 235.94          & -0.17 & 90.37          & +1.19  & 270.37          & +0.58 & 55.68          & +1.78  & 304.09          & +1.95\\
GRA &  124.53          & +0.12  & 235.94          & -0.18 & 89.88          & +0.78  & 269.88          & +0.98 & 55.20          & +1.44  & 304.27          & +1.82\\
SAV &  124.58          & +0.13  & 235.90          & -0.18 & 89.91          & +0.81  & 269.49          & +1.32 & 55.37          & +1.59  & 303.52          & +2.41\\
\hline
CDG &  123.89          & -0.14  & 235.46          & +0.28 & 89.31          & +0.32  & 269.33          & +1.46 & 54.05          & +0.48  & 303.28          & +2.40\\
\hline
\multicolumn{13}{l}{\footnotesize Notes: WS=Winter Solstice, SS=Summer Solstice, E=Equinox. GAL=Galleriano, GRA=Gradisca, SAV=Savalons, CDG=Castello di Godego.}\\
\end{tabular}
}
\end{table*}

\begin{table*}
\caption{\label{tab:lss} Refraction and horizon altitude corrected
  azimuths for moon rising (R) and setting (S) at major ($\pm(\varepsilon+i)$) and minor ($\pm(\varepsilon-i)$) standstills for 1500 B.C..}  
\tabcolsep 1.8mm \centerline{
\begin{tabular}{ccccccccc}
\hline
Site & \multicolumn{2}{c}{$-(\varepsilon+i)$} & \multicolumn{2}{c}{$-(\varepsilon-i)$} & \multicolumn{2}{c}{$\varepsilon-i$} & \multicolumn{2}{c}{$\varepsilon+i$}\\
\hline
    & R               & S               & R               & S               &  R              & S               & R               & S \\
\hline
GAL & 133.26$\pm$0.29 & 227.00$\pm$0.31 & 117.06$\pm$0.23 & 243.46$\pm$0.25 &  63.50$\pm$0.18 & 295.90$\pm$0.16 &  47.54$\pm$0.20 & 311.89$\pm$0.19\\ 
GRA & 133.18$\pm$0.30 & 226.96$\pm$0.30 & 116.83$\pm$0.24 & 243.48$\pm$0.25 &  63.67$\pm$0.17 & 295.00$\pm$0.15 &  47.61$\pm$0.20 & 311.34$\pm$0.18\\ 
SAV & 133.63$\pm$0.28 & 227.02$\pm$0.31 & 117.09$\pm$0.23 & 243.45$\pm$0.25 &  64.34$\pm$0.16 & 295.88$\pm$0.16 &  48.04$\pm$0.19 & 311.47$\pm$0.18\\ 
\hline
CDG & 132.76$\pm$0.30 & 226.25$\pm$0.26 & 116.39$\pm$0.24 & 243.13$\pm$0.23 &  62.16$\pm$0.23 & 295.08$\pm$0.15 &  46.84$\pm$0.23 & 310.95$\pm$0.18\\ 
\hline
\end{tabular}
}
\end{table*}

\section{\label{sec:orient}Orientation analysis}

At variance with the case of direct azimuth field measurements, the
availability of digital elevation data allows one to determine the
orientation of a rectilinear structure by means of simple least
squares fitting. This implicitly takes into account the
three-dimensional shape of the earthen structure. For this purpose, we
devised a software procedure that determines the centroids of a number
of cross sections along an interactively selected portion of the
embankment and fits a straight line to their x and y coordinates,
deriving the azimuth and its associated statistical error.

AR86 assumed that the horizon altitude was close to zero; however, the
Alps are clearly visible from the sites. In the direction of the
highest peaks (2500-3000 m) the horizon altitude exceeds 2 degrees,
hence requiring a significant correction to the derived
declinations. In this study the horizon profiles were computed using
the procedure described in Patat (\cite{patat11}), which includes
terrestrial refraction and Earth's curvature corrections. The digital
elevation data adopted for the calculations are those provided by the
Shuttle Radar Topographic Mission (SRTM; Farr et al. \cite{farr}),
which has a resolution of 90 m. A similar approach to horizon
calculation was adopted by Pimenta, Tirapicos \& Smith
(\cite{pimenta}) in their analysis of megalithic enclosures.
The resulting horizon profiles are shown in Figure~\ref{fig:horprof}.

Given the location of the three sites, the natural horizon is mostly
placed at distances larger than 30 km and, therefore, the expected
accuracy on the computed altitude is better than 0.1 degrees (Patat
\cite{patat11}). We note that the values corresponding to sharp
mountain ridges are systematically smaller than real, because of the
smoothing introduced by the 90 m SRTM resolution bins. A direct
altitude measurement of the peak of M. Canin (2587 m), visible from
all the three sites and placed at a distance of about 50 km, gave a
value of 2.50 degrees. This is to be compared to the 2.44 degrees
estimated by the numerical procedure, which gives an idea of the
magnitude of the systematics affecting the horizon calculations.

An example horizon computation is presented in Fig.~\ref{fig:horizon}
for the site of Gradisca. The calculations were done assuming an eye
position of 5.5 meters above the surrounding plain (which is about 1.5
meters above the top of the embankments).  In the azimuth
range 148-256 degrees (i.e. in the direction of the Adriatic sea) the
horizon altitude is negative. This is because the Friulian plain
steadily declines to the sea level with a gradient of $\sim$5.5 m
km$^{-1}$ roughly aligned with the N-S direction (see also
Sect.~\ref{sec:disc}).

There is a caveat one should always bear in mind when considering
risings/settings on the natural horizon. The procedure described above
implicitly assumes that this is actually visible from the
sites. Although the elevation of the embankments ($>$4 m) positions an
hypothetical observer at more than 5.5 meters above ground, high trees
can completely cover significant portions of the horizon.  For
instance, a 25 m high tree placed at 1.2 km prevents the horizon
visibility below $\sim$1 degree. In other words, horizon visibility
requires that the area surrounding the sites was free of high
vegetation within a radius of a couple of km. This is by no means
guaranteed and it constitutes a serious issue, which is often
neglected in orientation analyses published in the literature (see for
instance AR86).

The availability of a digital horizon profile enables an easy
numerical derivation of declinations ($\delta$) from measured azimuths
($a$) and vice-versa. The procedure (which includes astronomical
refraction correction and statistical error estimate) is described in
Appendix~\ref{sec:appa}.  Using this approach, we computed the
apparent azimuths for the sunrise (SR) and sunset (SS) at the Winters
Solstice (WS), at the Equinoxes (E) and at the Summer Solstices
(SS). The results, calculated for 1500 B.C. (obliquity
$\varepsilon$=23.87 degrees), are presented in Table~\ref{tab:sun},
which includes the horizon altitude (indicated by $h$).  These are
used as terms of reference in the orientation analysis that is
presented in the following subsections.  The table includes also the
relevant data for the site of Castello di Godego (Bianchin Citton
\cite {bc89}), which will be discussed in Sect.~\ref{sec:disc}. All
calculations were done for the Sun center and for the lower and upper
limbs of the solar disc. The azimuth range corresponding to the upper
and lower limbs is about $\pm$0.4 degrees from the sun center values.

Since major and minor lunar standstills are listed as possible target
alignments by AR86 for Galleriano and Gradisca (see their Table~2), we
performed a similar calculation for the relevant moon azimuths. The
results are presented in Table~\ref{tab:lss} (average lunar orbit
inclination $i$=5.14 degrees).  We emphasize that this table was
included only for the purposes of the comparison with the work of
AR86.

\subsection{\label{sec:gall}Galleriano}

\begin{figure*}
\centering
\includegraphics[width=17cm]{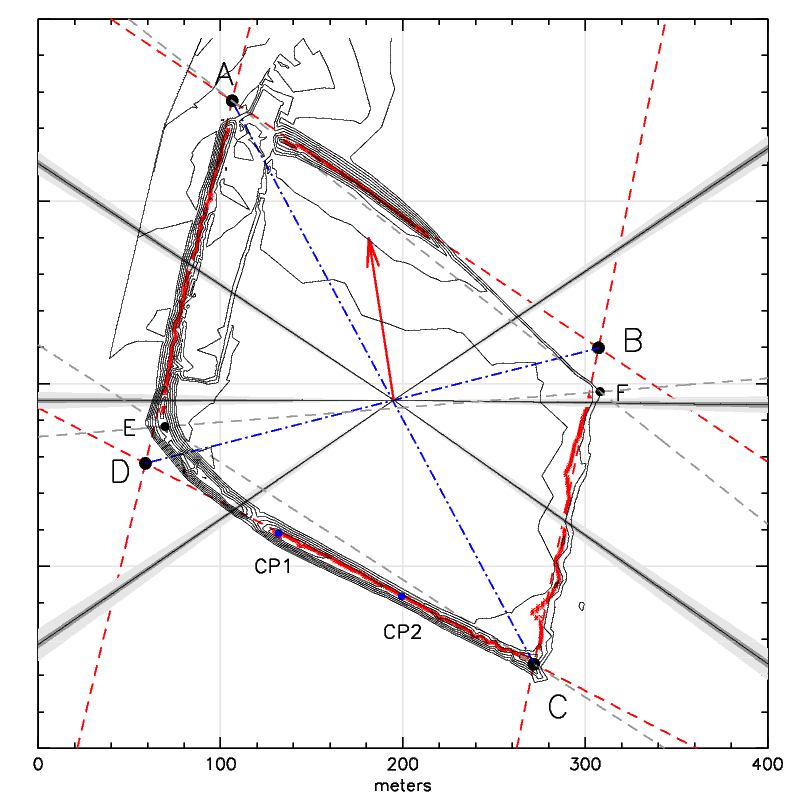}  
\caption{\label{fig:gall} DEM contour plot for Galleriano. Contours
  are drawn with a step of 0.5 m. The dashed lines trace the best fit
  directions computed for the points marked in red.  The points
  identified as A, B, C, and D are the intersections between the best
  fit lines.  The dotted-dashed lines trace the corresponding
  diagonals. E and F mark the western and eastern corners,
  respectively. The light-colored dashed line traces the direction
  defined by these two points. CP1 and CP2 indicate the positions of
  the two check points used to validate the DEM absolute
  orientation. The light-shaded areas indicate the reference
  astronomical directions listed in Table~\ref{tab:sun} ($\pm$5 SD;
  the dark-shaded areas trace the lower-upper limb ranges). The arrow
  in the center indicates the direction of maximum gradient (see
  Sect.~\ref{sec:disc}).The grid is oriented along the true N-S and
  E-W directions (N to the top and E to the right, respectively).}
\end{figure*}

The DEM contour plot for Galleriano is presented in
Fig.~\ref{fig:gall}. The figure shows the best fit directions for the
four sides. Only the data from the rectilinear portions of the
embankment were used in the least squares fitting. The points marked
as A, B, C and D indicate the geometrical intersections between the
best fit lines. With the only exception of C (southern corner), they
do not coincide with the physical corners of the earthwork, but they
provide non-subjective points for defining the diagonals. For the sake
of completeness, we also marked the western and eastern corners E and
F. While the position of E is well defined, F is much less so, because
of the poor state of the AB and CB sides. For the same reason, the
directions defined by these two sides are very uncertain.  The figure
also displays the reference solar directions listed in
Table~\ref{tab:sun} with the $\pm$5 standard deviation (SD) confidence
levels (light gray shaded areas) and the lower-upper solar limb
interval (dark gray shaded areas).

\begin{table}
\caption{\label{tab:gall} Orientation data for Galleriano for the
  eastern (upper table) and western (lower table) directions.}
\tabcolsep 1.75mm \centerline{
\begin{tabular}{cccccc}
\hline
        &            &         &               & \multicolumn{2}{c}{AR86} \\
\cline{5-6} 
ID      &   $a$      & $h$     & $\delta$      & $a$ & $\delta$ \\ 
\hline
AB      &  123.99$\pm$0.22 &+0.20 & -23.19$\pm$0.20 &   -    &   -  \\
CB      & \p11.55$\pm$4.77 &+2.31 & +44.86$\pm$0.77 & \p10.2 & +42.5\\
DC      &  117.33$\pm$0.06 &+0.21 & -19.17$\pm$0.15 & 119.7  & -20.6\\
DA      & \p13.46$\pm$1.54 &+2.20 & +44.33$\pm$0.40 & \p19.7 & +40.3\\
\hline
AC      &  151.82$\pm$0.65 &-0.21 & -38.62$\pm$0.34 & -      & -\\
DB      & \p75.70$\pm$1.30 &+1.00 & +10.23$\pm$0.92 & \p77.8 & \p+8.0\\
\hline
AF      & 128.3$\pm$0.4    &+0.16 & -25.86$\pm$0.28 & 135.7  &  -30.3\\
EC      & 122.7$\pm$0.4    &+0.12 & -22.44$\pm$0.32 & -      &-\\
EF      &\p85.4$\pm$0.5    &+0.98 & +3.59$\pm$0.33  & -      &-\\
\hline
        &                  &                        &        & \\
\hline
BA      &  303.99$\pm$0.22 &+1.97 & +24.10$\pm$0.18 &   -    &   -  \\
BC      &  191.55$\pm$4.77 &-0.21 & -43.64$\pm$0.97 & 190.2  & -42.5\\
CD      &  297.33$\pm$0.06 &+1.54 & +19.17$\pm$0.12 & 299.7  &+20.6\\
AD      &  193.46$\pm$1.54 &-0.20 & -43.40$\pm$0.41 & 199.7  & -40.3\\
\hline
CA      &  331.82$\pm$0.65 &+1.98 & +39.25$\pm$0.39 & -      & -\\
BD      &  255.70$\pm$1.30 &-0.10 & -10.44$\pm$0.94 & 247.8  & \p-8.0\\
\hline
FA      & 308.3$\pm$0.4    &+2.02 & +26.83$\pm$0.22 & 315.7  & +30.3\\
CE      & 302.7$\pm$0.4    &+1.72 & +23.06$\pm$0.37 & -      &-\\
FE      & 265.4$\pm$0.5    &+0.50 & -3.24$\pm$0.39  & -      &-\\
\hline
\end{tabular}
}
\end{table}

The best fit results are reported in Table~\ref{tab:gall}, which lists
azimuth ($a$), horizon altitude ($h$), corrected declination
($\delta$) and their associated errors (SD) for each alignment and for
both rising and setting directions. For convenience the table also
lists the results by AR86 (see their Table 2), who considered ``sides,
where measurable, as well as any visible diagonal connecting definable
corners of the monument''. For this reason our cross-matching in
Table~\ref{tab:gall} is tentative.

\begin{figure*}
\centering
\includegraphics[width=17cm]{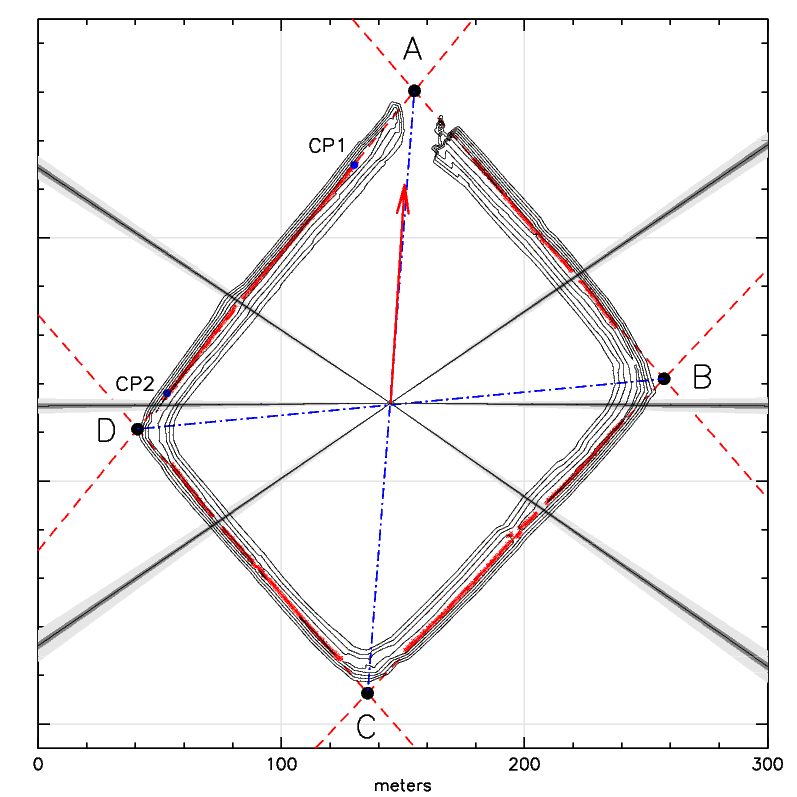}  
\caption{\label{fig:grad} Same as Fig.~\ref{fig:gall} for Gradisca.}
\end{figure*}

In the comparison that follows we neglect the discrepancies on the
directions of sides CB and DA, because they are of no obvious
astronomical interest.

The best preserved side of the earthwork is the one we designate as CD
(SW). Although it shows a marked bending approaching the western
corner, it is characterized by a rectilinear segment that extends for
about 160 m. For the validation of the DEM orientation we performed a
direct field measurement using two check points (marked as CP1 and CP2
in Fig.~\ref{fig:gall}) placed 75.8 m apart. The true azimuth,
averaged over twelve sun fixes, is 117.19$\pm$0.01 degrees. This is
fully consistent with the value derived from the DEM best fitting
(117.33$\pm$0.06 degrees). The obvious match is with the value 119.7
reported by AR86 for one of the two long sides (the other being 135.7
degrees). The discrepancy is about 2.4 degrees and it is very
difficult to explain in terms of measurement errors. A possible
explanation is that AR86 used a different set of reference points. In
this respect, we note that the azimuth identified by points E and C
(see Fig.~\ref{fig:gall}) is 122.7$\pm$0.4 degrees, so that the value
reported by AR86 appears to be very close to the average of the
two. This suggest that they may have placed one of the two reference
points not on the physical western corner E, but somewhere on the
bended part of the SW side.

As far as the other long side AB is concerned, we remark that the
apparent statistical azimuth accuracy ($\pm$0.22 degrees) only refers
to the portion that we used (Fig.~\ref{fig:gall}). Although some
traces of the embankment are visible all the way to the eastern
corner, with the currently available data is not possible to tell the
exact shape and direction of the missing portion.  The best fit
azimuth of the surviving part (123.99$\pm$0.22 degrees) deviates
significantly from the value reported by AR86 (135.7 degrees), clearly
indicating that their measurement refers to a different alignment. The
most natural alternative is that identified by the northern and
eastern corners (A and F, respectively); however, the resulting
azimuth (128.3$\pm$0.4 degrees) deviates by 7.4 degrees from the AR86
value. We could not identify any obvious feature with the AR86 azimuth
in our DEM. This clearly illustrates the difficulties inherent to the
study of sites like Galleriano, where the identification of possible
alignments is rather subjective. Based on the observed discrepancies,
we argue that the alignments considered by AR86 for sides AB and CD
(lunar standstills rise and setting) are strongly debatable. In all
fairness, our new DC azimuth determination (117.33$\pm$0.06) is
consistent to within $\sim$1 sigma with the moon rising at
$-(\varepsilon-i)$ (117.06$\pm$0.23 degrees; see
Table~\ref{tab:lss}). Although this is a statistically significant
match, given the lack of similar alignments in the other sites studied
in this work (see next sections and the discussion), we attribute it
to a chance coincidence.

The comparison to the solar reference directions suggests a possible
alignment of the long sides to the WSSR and SSSS. Nevertheless, the
CE direction differs from the SSSS by 0.95$\pm$0.43 degrees, which
is a deviation at the 2.2-sigma level. The difference with the WSSR is
larger (2.5$\pm$0.5 degrees), making the alignment statistically unlikely.
Because of the poor state of the AB side, we reckon it is not possible
to conclude whether or not this reflects a solstitial alignment, as
the measured values may suggest.

As for the major/minor diagonals, we note that both deviate from the
NS/EW, hence contradicting the common statement about cardinal
alignments as a characteristic feature of these sites. Based on the
measured azimuth values, we can only conclude that the corners of the
earthwork are very roughly oriented to the cardinal points. AR86
report an azimuth of 77.8 for the minor diagonal. This value is very
different from that computed for the EF corners direction
(85.4$\pm$0.5) suggesting that, again, they adopted a different pair
of points. The AR86 value can be obtained if one of the two points is
placed in F and the other on the DC side, where this starts to bend in
the direction of E. As this choice appears to be completely arbitrary,
we believe no information can be extracted from the derived azimuth.

Given the state of the embankment around the eastern corner, it is not
possible to firmly establish its position with sufficient
accuracy. Nevertheless, based on the existing data, we reckon an
equinoctial alignment can be excluded.

\subsection{\label{sec:grad}Gradisca}

The DEM contour plot for Gradisca is presented in
Fig.~\ref{fig:grad}. Points and lines have the same meaning as in
Fig.~\ref{fig:gall} (see also the previous Sect.~\ref{sec:gall}). The
best fit results are presented in Table~\ref{tab:grad}.

The sides of this site are well defined and very rectilinear, allowing
an accurate determination of their azimuths. The corners are also well
defined, with the only exception of the northern one, which was
demolished at the beginning of the 20th century to give access to
cultivations. As in the previous case, in the analysis we use the best
fit lines intersection to compute the corners positions in an
objective way. The fitted azimuths are listed in Table~\ref{tab:grad},
where they are compared to the previous estimates by AR86.

\begin{table}
\caption{\label{tab:grad} Orientation data for Gradisca for the
  eastern (upper table) and western (lower table) directions.}
\tabcolsep 1.75mm \centerline{
\begin{tabular}{cccccc}
\hline
        &            &         &               & \multicolumn{2}{c}{AR86} \\
\cline{5-6} 
ID      &   $a$      & $h$     & $\delta$      & $a$ & $\delta$ \\ 
\hline
AB      &  139.08$\pm$0.15 &-0.18  & -32.41$\pm$0.20  & 140.5  & -32.9 \\
CB      & \p43.33$\pm$0.10 &+2.22  & +31.92$\pm$0.13  & \p42.9 & +30.0 \\
DC      &  138.94$\pm$0.12 &-0.17  & -32.33$\pm$0.19  & 139.7  & -32.6 \\
DA      & \p39.31$\pm$0.12 &+2.12  & +34.01$\pm$0.15  & \p41.0 & +31.0 \\
\hline
AC      &  184.45$\pm$0.09 &-0.21  & -44.76$\pm$0.22  & -      & -     \\
DB      & \p84.55$\pm$0.13 &+0.60  &\p+3.83$\pm$0.16  & \p85.0 & \p+3.0\\
\hline
        &                  &      &                   &        &       \\
\hline
BA      &  319.08$\pm$0.15 &+2.60 & +33.58$\pm$0.12   & 320.5  & +32.9 \\
BC      &  223.33$\pm$0.10 &-0.19 & -31.10$\pm$0.19   & 222.9  & -30.0 \\
CD      &  318.94$\pm$0.12 &+2.63 & +33.53$\pm$0.12   & 319.7  & +32.6 \\
AD      &  219.31$\pm$0.12 &-0.20 & -33.29$\pm$0.20   & 221.0  & -31.0 \\
\hline
CA      &\p\p4.45$\pm$0.09 &+2.45 & +45.95$\pm$0.14   & -      & -     \\
BD      &  264.55$\pm$0.13 &+0.84 &\p-3.54$\pm$0.16   & 265.0 & \p-3.0\\
\hline
\end{tabular}
}
\end{table}

The DEM orientation was validated using two check points (marked as
CP1 and CP2 in Fig.~\ref{fig:grad}) placed 130 m apart on the NW
side. The azimuth, averaged over ten sun fixes, is 39.22$\pm$0.01
degrees, which is in good agreement with the best fit value
(39.31$\pm$0.12 degrees).  For this side, AR86 report an azimuth of 41.0
degrees, which is about 0.7 degrees off with respect to our
measurements. Similar deviations are seen in the other angles. Given
the very good conditions of the earthwork, we deem the mismatch is due
to the subjective positioning of the reference point pairs on the
corners defining the relevant directions in the AR86 study.

While the sides AB and BC are almost parallel (the azimuth difference
is 0.14$\pm$0.19), the azimuths of BC and DA differ by 4.02$\pm$0.15
degrees. The angles formed by the diagonals differ by 9.9$\pm$0.2
degrees from the right angle. A similar deviation from orthogonality
is seen in adjacent sides.

A quick comparison to the solar reference directions shows that none
of the alignments is statistically significant (see also
Fig.~\ref{fig:grad}). The closets match is that of the minor diagonal
DB, which is $\sim$5.9 degrees off (i.e. more than 30 SD). A similar
deviation from the NS direction is displayed by the major diagonal AC.
This rough alignment to the NS and EW directions is probably the
origin of the diffused perception that these sites have a cardinal
orientation. Although the case of Gradisca may be regarded as more
accurate than Galleriano (see Sect.~\ref{sec:gall}), we reckon the
deviations are sufficiently large to make the match to astronomically
relevant directions statistically unlikely.

As for the case of Galleriano, AR86 considered minor and major lunar
standstills for the sides CB and DA. The same matches were considered
for sides AB and DC, but they were listed with question marks (see
their Table 2). Our azimuth measurements and horizon corrections
produce declinations that deviate in a statistically very significant
way from the nominal lunar azimuths (see Table~\ref{tab:lss}). In this
respect, it is important to emphasize that AR86 commented that the
alignments to lunar extrema were only considered to pragmatically
follow Thom's paradigm (see their note 8). No explicit claim was made
about their significance.

\begin{figure*}
\centering
\includegraphics[width=17cm]{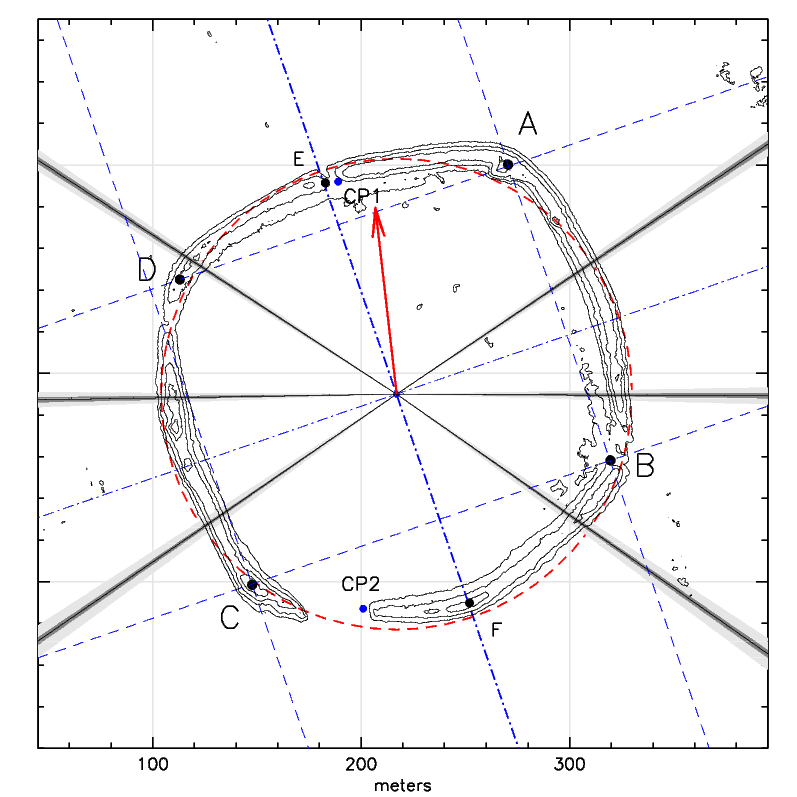}  
\caption{\label{fig:sava} Same as Fig.~\ref{fig:gall} for Savalons.}
\end{figure*}

\subsection{\label{sec:sava}Savalons}

The DEM contour plot for Savalons is presented in
Fig.~\ref{fig:sava}. The orientation was validated using two
check-points placed 205 meters apart at the NW and SW openings (marked
as CP1 and CP2, respectively). The derived azimuth is 176.7$\pm$0.1
degrees, which nicely matches the value measured on the DEM
(176.6$\pm$0.3 degrees). As anticipated, this site differs
substantially from the other two, in that it does not have a
clear quadrangular shape. Although four corners (marked as A, B, C and
D in Fig.~\ref{fig:sava}) can be identified, the sides present marked
bendings alternated by rectilinear segments. This is particularly
evident in the southern side of the embankment.

\begin{table}
\caption{\label{tab:sava} Orientation data for Savalons for the
  eastern (upper table) and western (lower table) directions.}
\tabcolsep 4.5mm \centerline{
\begin{tabular}{cccc}
\hline
ID      &   $a$      & $h$     & $\delta$      \\ 
\hline
AB      &  160.87$\pm$0.34 & -0.30 & -41.94$\pm$0.24\\
CB      & \p70.84$\pm$0.28 & +1.43 & +13.90$\pm$0.22\\
DC      &  166.69$\pm$0.38 & -0.29 & -43.49$\pm$0.24\\
DA      & \p70.72$\pm$0.28 & +1.42 & +13.98$\pm$0.21\\
\hline
AC      &  211.34$\pm$0.31 & -0.28 & -37.26$\pm$0.24\\
DB      &  112.78$\pm$0.24 & +0.20 & -15.83$\pm$0.26\\
\hline
        &                  &       &                \\
\hline
BA      &  340.87$\pm$0.34 & +2.48 & +43.08$\pm$0.16\\
BC      &  250.84$\pm$0.28 & -0.01 & -13.67$\pm$0.24\\
CD      &  346.69$\pm$0.38 & +2.73 & +44.86$\pm$0.21\\
AD      &  250.72$\pm$0.28 & -0.01 & -13.75$\pm$0.24\\
\hline
CA      & \p31.34$\pm$0.31 & +2.61 & +38.45$\pm$0.19\\
BD      &  292.78$\pm$0.24 & +1.62 & +16.51$\pm$0.16\\
\hline
\end{tabular}
}
\end{table}

The earthwork appears to be roughly inscribed in a circle with a
diameter of about 226 meters (dashed circle) and is characterized by a
rough symmetry axis passing through the center of this circle. The
position of this axis can be estimated using a numerical procedure
based on the available DEM data. Very briefly, the best fit axis is
found numerically, with a brute-force procedure that maximizes the
match between the collection of the original embankment centroids and
their axially symmetric images. The resulting azimuth is 161.0$\pm$0.4
degrees.  For convenience, we marked with E and F the positions where
this axis intersects the highest points on the embankment.

An independent estimate can be obtained considering the azimuths of
the segments AB, CB, DC and DA. These are reported in
Table~\ref{tab:sava}, together with the horizon altitude and the
implied refraction corrected declination. Excluding the deviant case
of DC, the average azimuth is 160.81$\pm$0.18 degrees, which is fully
consistent with the above value\footnote{In this calculation the
  directions perpendicular to CB and DA were considered.}. We note that
  the angles $\widehat{DAB}$ and $\widehat{ABC}$ differ by less than
  0.2 degrees from the right angle, so that DA and CB are
  parallel to within $\sim$0.1 degrees.

Like in the other two cases, we considered the directions defined by
the two diagonals AC and DB. Their azimuths are shown in
Table~\ref{tab:sava}. A quick inspection reveals that there is no
match with any of the reference values listed in Tables~\ref{tab:sun}
and \ref{tab:lss}. Interestingly, this holds even considering the
directions defined by segments like FB, FA, FD and so on. We therefore
conclude that there is no astronomically significant alignment for
this site.

\section{\label{sec:disc}Discussion and conclusions}

\begin{figure*}
\centering \includegraphics[width=14.0cm]{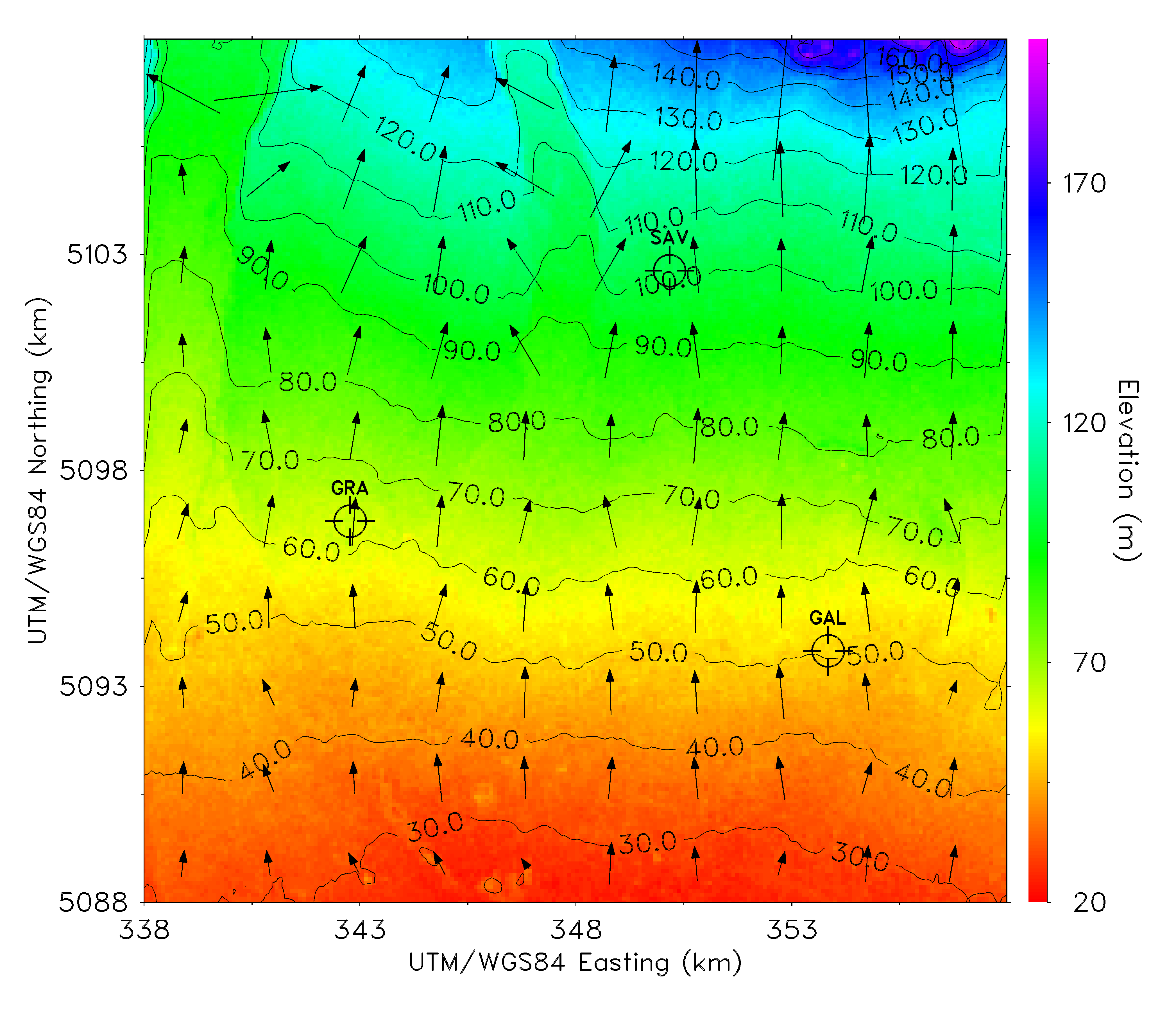}
\caption{\label{fig:gradient} Gradient vector field for the Friulian
  high-plain. The arrows indicate the direction of maximum slope;
  vector lengths are proportional to the gradient. The positions of
  the three sites are marked.}
\end{figure*}

Although they are all placed within a radius of about 5 km, the three
sites examined in this paper do not show common orientations. In
addition, none of the sites presents statistically significant
alignments to relevant solar or lunar directions. These facts imply
that these earthworks, at least in their current shape, were not
erected following astronomical criteria, which would naturally provide
a unifying schema. The most likely conclusion is that the orientation
of the sites was dictated by other considerations and constraints,
plausibly related to the local geomorphological properties of the
environment.

In her review of the Friulian castellieri, C\`assola Guida
(\cite{cg80}, p. 16) discussed the orientation aspects of the three
sites examined in this paper, commenting that they all share a common
feature, i.e. a rough alignment of their corners along the NS, EW
direction. C\`assola Guida (\cite{cg80}) also notes that in the
Friulian plain all rivers and water streams flow in the NS direction
and, therefore, the observed orientations may have been designed to
guarantee proper draining, especially in the case of floods. The
deviations from the underlying criterion are attributed to local
morphological variations.

For assessing this idea in a more quantitative way, we derived the
gradient map of the high Friulian plain using the SRTM data (90 m
resolution). The gradient was computed as the maximum slope of best
fit planes on a 2$\times$2 km$^2$ ($\sim$490 SRTM pixels) grid. The
result is presented in Fig.~\ref{fig:gradient}, which displays also
the position of the three sites. Although small variations are seen,
the average gradient is definitely aligned with the NS direction. The
largest deviations are seen to the north (due to the rising of a
glacially formed moraine) and to the west (caused by the depression
associated to the Tagliamento river bed).  The local gradients (in m
km$^{-1}$) and gradient azimuths (in decimal degrees), calculated
within squares of 2$\times$2 km$^2$ and 1$\times$1 km$^2$ centered on
the sites, are presented in Table~\ref{tab:gradient}. The gradient
azimuths (2$\times$2 km$^2$) are indicated with an arrow in
Figs.~\ref{fig:gall}, \ref{fig:grad} and \ref{fig:sava}.

Although the SRTM resolution is not sufficient to establish the local
slopes at scales below $\sim$1 km, it is clear that the gradient
azimuths at the three sites are very similar\footnote{It is worth
  noting that, on such small scales, the local gradient is much more
  subject to human activities than on large scales, and what is
  measured today may differ substantially from the values at
  construction time.}. The observed variations are much smaller than
those seen, for instance, in the azimuths of the DC sides (see
Figs.~\ref{fig:gall}, \ref{fig:grad} and \ref{fig:sava}). One common
feature between the sites is that none of their sides is perpendicular
to the gradient direction (a possible exception is the portions of the
Savalons embankment indicated as CF in Fig.~\ref{fig:sava}, but this
is anyway downstream). This suggests that, indeed, water draining may
have been a relevant aspect while planning the earthworks.

\begin{table}
\caption{\label{tab:gradient} Local gradients (m km$^{-1}$) and azimuths (degrees).}
\centerline{
\tabcolsep 2mm
\begin{tabular}{lccccc}
\hline
           & \multicolumn{2}{c}{2$\times$2} & &\multicolumn{2}{c}{1$\times$1} \\
\cline{2-3} \cline{5-6}
Galleriano & 5.1$\pm$0.1 & -8.8$\pm$1.2 & & 5.3$\pm$0.3 & +5.2$\pm$3.3\\
Gradisca   & 5.4$\pm$0.1 & +3.7$\pm$1.3 & & 4.0$\pm$0.4 & +2.3$\pm$5.6\\
Savalons   & 6.0$\pm$0.1 & -6.5$\pm$1.0 & & 5.1$\pm$0.3 & +0.1$\pm$3.6\\
\hline
\end{tabular}
}
\end{table}

In this respect it is important to note that the azimuth of the {\it
  cardines} in the Roman centuriation of this area (the {\it ager} of
Aquileia, about 35 km to the south) is $\sim$158 degrees (Prenc
\cite{prenc02}). This corresponds to an inclination of $\sim$22
degrees with respect to the NS direction (and hence with the average
gradient direction). This ensured that both {\it cardines} and {\it decumani}
(forming the orthogonal centuriation grid) would have significant
slopes for an efficient water drainage. This azimuth is close to that
determined here for the symmetry axis of Savalons, and not too
different (in terms of resulting slope) from the AB and DC sides of
Galleriano and Gradisca.

Because of some similarities shared with the Friulian earthworks and
the attention it received in the astronomical context (see for
instance Romano \cite{romano}), it is worth discussing here the case
of Castello di Godego. This earthwork is located in the neighboring
Veneto region (longitude 11.8654E, latitude 45.6701N; see
Fig.~\ref{fig:map}) and it is dated back to the Bronze Age (Bianchin
Citton \cite{bc89}). The shape of the embankment is quadrangular, with
sides of 200-230 meters. The corners are roughly aligned with the
cardinal directions. Romano (\cite{romano81}) studied its orientation,
concluding that the sides point to the sun rise and setting at the
solstices, while the major diagonal is well aligned to the EW
direction. The results of that study are also included in AR86.  These
claims were debated by archaeologist Bianchin Citton (\cite{bc91}),
who excluded that astronomical criteria were prioritary in the
construction of this and other similar sites\footnote{The Friulian
  castellieri are cited by Bianchin Citton (\cite{bc91}) in this
  context, although no mention is made to any specific site.}.  The
findings of Romano (\cite{romano81}) were questioned based on a number
of arguments by Lupato (\cite{lupato00}), to which we refer the reader
for a detailed account. Here we will restrict the discussion to the
comparison with the Friulian sites presented in the article and its
implications on our conclusions.

\begin{figure*}
\centering
\includegraphics[width=17cm]{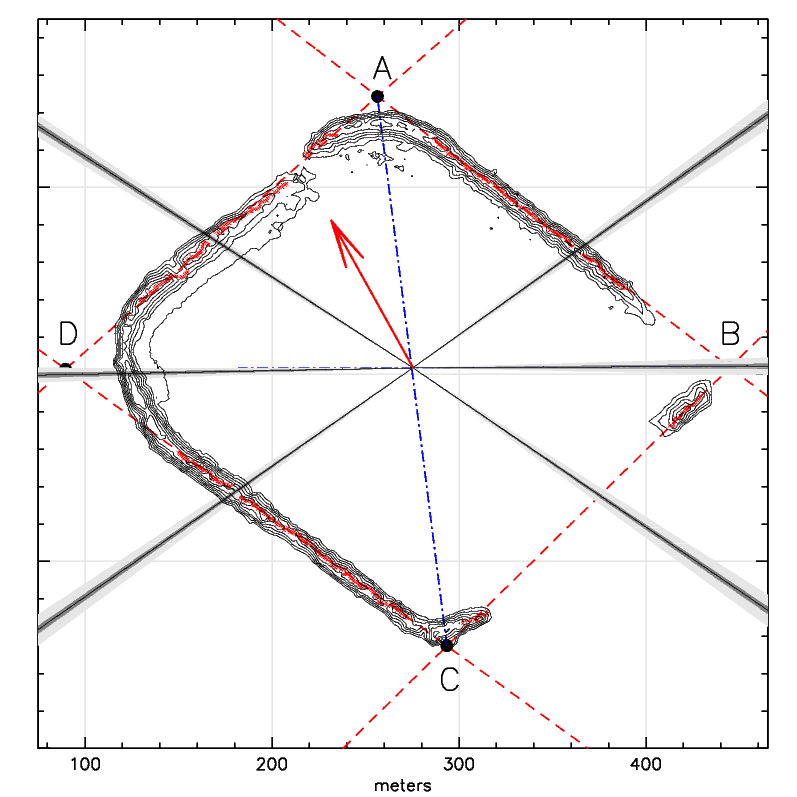}  
\caption{\label{fig:cdg} Same as Fig.~\ref{fig:gall} for Castello di
  Godego. The digital data used to produce the contour plot were
  obtained from the HELICA archive. The arrow in the center indicates
  the direction of maximum gradient (see text).}
\end{figure*}

To ensure a homogeneous analysis, we retrieved the LIDAR digital
elevation data from the HELICA digital archive (resolution 1m) and we
run the same numerical procedures used for the other sites. The DEM 
is presented in Fig.~\ref{fig:map_cdg} and the
contour plot is shown in Fig.~\ref{fig:cdg}, together with the reference sun
directions (see Table~\ref{tab:sun}, last row) and the best fit
directions. These are listed in Table~\ref{tab:cdg}, together with the
horizon and refraction corrected declinations. For comparison, the
AR86 data are also included.

\begin{table}
\caption{\label{tab:cdg} Orientation data for Castello di Godego for
  the eastern (upper table) and western (lower table) directions.}
\tabcolsep 1.3mm \centerline{
\begin{tabular}{cccccc}
\hline
        &            &         &               & \multicolumn{2}{c}{AR86} \\
\cline{5-6} 
ID      &   $a$      & $h$     & $\delta$      & $a$ & $\delta$ \\ 
\hline
AB      &  127.56$\pm$0.17 &-0.15  & -25.89$\pm$0.20  & 124.0$\pm$1.0  & -23.5 \\
CB      & \p45.47$\pm$0.33 &+1.16  & +29.91$\pm$0.25  &  -             &  -  \\
DC      &  125.97$\pm$0.15 &-0.13  & -24.88$\pm$0.19  & 125.0$\pm$1.0  & -24.1\\
DA      & \p48.85$\pm$0.45 &+0.76  & +27.57$\pm$0.36  & \p54.0$\pm$1.5 & +23.7\\
\hline
AC      &  172.80$\pm$0.44 &-0.16  & -44.75$\pm$0.16  & -              &     - \\
DB      & \p89.86$\pm$0.31 &+0.32  &\p-0.10$\pm$0.26  & \p90.0$\pm$2.5 & \p\p0.0\\
\hline
        &                  &       &                  &                &    \\
\hline
BA      &  307.56$\pm$0.17 &+2.53  & +26.95$\pm$0.14  & 304.00$\pm$1.0 & +23.5\\
BC      &  225.47$\pm$0.33 &+0.46  & -29.43$\pm$0.23  &  -             &  -   \\
CD      &  305.97$\pm$0.15 &+2.42  & +25.87$\pm$0.14  & 305.0$\pm$1.0  & +24.1\\
AD      &  228.85$\pm$0.45 &+0.39  & -27.51$\pm$0.31  & 234.0$\pm$1.5  & -23.7\\
\hline
CA      &  352.80$\pm$0.44 &+3.56  & +47.10$\pm$0.15  & -              & - \\
BD      &  269.86$\pm$0.31 &+1.51  &\p+0.69$\pm$0.26  & 270.0$\pm$2.5  & \p\p0.0 \\
\hline
\multicolumn{6}{c}{Note: The errors of the AR86 azimuths are from Romano (\cite{romano81}).}
\end{tabular}
}
\end{table}

A coarse inspection to Fig.~\ref{fig:cdg} reveals that the SE side
(which we indicate as CB ) was almost completely erased\footnote{Note
  that the map orientation in Bianchin Citton (\cite{bc89}) is
  incorrect (see her Figs.~2 and 5).} and it is therefore very
difficult to study. In the best fit procedure we used the data of the
surviving portion close to the eastern corner and a small portion of
the southern corner. We remark that the result is unreliable and
cannot be used in the analysis; we list it only for the sake of
completeness. The case of the NW side (here indicated as DA) is also
rather problematic, because the embankment was significantly modified
in the portion close to the northern corner. On the contrary, the
status of the other two sides (AB and DC) is reasonably good and the
azimuth estimates are fairly accurate.

For these two sides, the AR86 azimuths differ from ours by $\sim$3.6
degrees (AB) and $\sim$1.0 degrees (CD), and this translates into
significant declination discrepancies. In particular, our values are
statistically inconsistent with solstitial alignments. The deviation
is even more significant for the sun setting directions, especially
for the SSSS (CD and BA), for which the horizon altitude (neglected in
AR86) is substantial (see also the discussion in Lupato
\cite{lupato00}). The AC diagonal deviates from the NS direction by
7.2$\pm$0.4 degrees, while the DB diagonal lies to within 0.1$\pm$0.3
degrees from the EW direction (similarly to what is reported by
AR86). However, the eastern corner is in an extremely bad shape. While
in our case this was determined as a best fit intersection position,
it is not entirely clear how this was achieved in the original work by
Romano (\cite{romano81}).

The present analysis confirms the approximate corner alignment to the
cardinal directions (see for instance Bianchin Citton \cite{bc89}),
which is more accurate than in the Friulian sites examined in this
study. On the other hand, the new azimuth determinations and horizon
corrections presented here do not confirm the accuracies claimed for
the alignments to the solar standstills.

Like in the case of the Friulian sites, it is important to consider
the orientation in the context of the local environment.  Balista
(\cite{balista}) reports an average slope of 5 meters km$^{-1}$ for
the area surrounding the earthwork, with a NW-SE gradient direction,
concluding that this is parallel to the long sides of the
earthwork. This fact was also noted by Lupato (\cite{lupato00}), who
argued that the orientation may be dictated by drainage
considerations\footnote{Lupato (\cite{lupato00}) also argued that the shape
of the embankment may have been substantially modified in modern times.
This matter needs to be settled on archaeological grounds.}.

In this respect we note that the short sides (BC and AD) are almost
perpendicular to the long sides (AB and CD). Therefore, if it is true
that aligning the long sides to the maximum gradient direction would
maximize the slope along them, it would almost minimize that along the
short ones. This would be at odds with the drainage hypothesis. For a
more quantitative study we ran a gradient analysis similar to the one
presented for the Friulian sites. The local gradient, computed in a 2$\times$2
km$^2$ area, is 4.2$\pm$0.2 m km$^{-1}$, with an azimuth of +151.0$\pm$2.2
degrees. This direction is indicated by the arrow in
Fig.~\ref{fig:cdg}. Clearly, the sides of Castello di Godego are all
significantly inclined with respect to the gradient direction reported
above, similarly to what is observed in the Friulian sites. This
strengthens the conclusion that the observed orientation may indeed
have been driven by practical needs related to the water flow.

One obvious limitation of the present work is that the sample of
castellieri we could examine is small. Future archaeological research
on the Friulian plain may enable a more statistically significant
analysis.  None the less, in the light of the above considerations, we
argue that a connection between the observed orientations and relevant
solar or lunar directions is highly improbable for these sites. The
lack of a consistent orientation pattern in the studied earthworks
strongly indicates that, at least in their current configuration, they
were not erected following common astronomical criteria.

\acknowledgements This paper is dedicated to the memory of G. Romano,
who first introduced F. P. to the history of astronomy.  The authors
are indebted to D. Plos, V. Dereani, C. Peloso and F. Tadina at HELICA
for the high resolution LIDAR survey performed on the Savalons site
and the valuable support given in the treatment of the data. The
authors are grateful to S. Marcuzzi for his kind help during the
preparation of the LIDAR survey, and to M. Calosi and D. Girelli for
their collaboration in the analysis of the total station survey
data. F.~P. wishes to thank profs. A. Aveni, E. Milone, C. Ruggles and
the late prof. G. Romano for their encouragement and useful
suggestions. The authors are also thankful to Dr. R.~D. Sampson, for
providing them with his atmospheric refraction measurements, and to
G. Lupato for the stimulating discussions on the site of Castello di
Godego. Data analysis and visualization were performed using software
written by the authors.

\newpage

\appendix

\begin{figure}
\centering
\includegraphics[width=8cm]{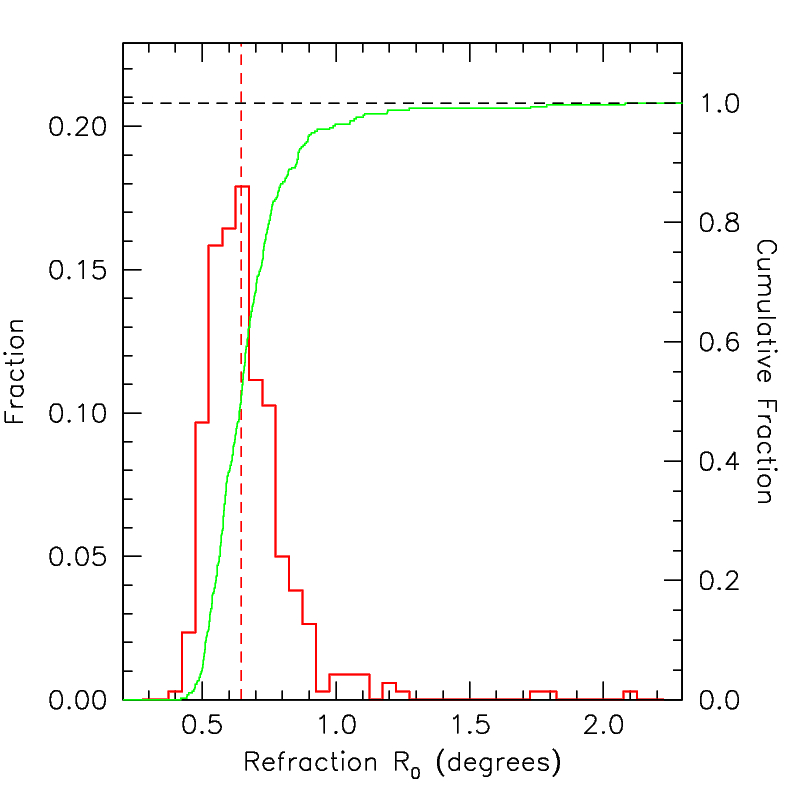}  
\caption{\label{fig:sampson} Refraction distribution adopted for the
  error estimates. The data are from Sampson et
  al. (\cite{sampson03}). The vertical dashed line marks the median
  value. The green curve is the distribution cumulative function
  (scale on the right vertical axis).}
\end{figure}

\section{\label{sec:appa} Appendix A - Horizon altitude, Refraction 
correction and error estimates}

The apparent declination $\delta^\prime$ of a celestial object
observed from a site at latitude $\phi$ can be calculated from its
azimuth $a$ and apparent altitude $h^\prime$ with the following
formula:

\begin{equation}
\label{eq:delta}
\delta^\prime = \arcsin \left ( \cos a \cos \phi \cos h^\prime + \sin \phi \sin
h^\prime \right )
\end{equation}

After indicating with $h$ the true altitude of the object, we
introduce the astronomical refraction $R$ defined as $R=h^\prime -
h$. One can demonstrate that the true declination can be derived
correcting the apparent declination given by Eq.~\ref{eq:delta} as
follows:

\begin{equation}
\delta = \delta^\prime - R \; \frac{\sin \phi - \sin \delta \sin (h^\prime-R)}
{\cos \delta \cos (h^\prime-R)}.
\end{equation}

This equation can be solved for $\delta$ analytically (considering
that $\sin\delta \approx \sin\delta^\prime$ and $\cos\delta \approx
\cos\delta^\prime$), or numerically (with an iterative process).  For
archaeoastronomical applications in which the horizon altitude
$h^\prime$ is only a few degrees (i.e. $h^\prime-R\approx$0), one can
derive the true declination via the following approximate expression:

\begin{equation}
\label{eq:delta2}
\delta \simeq \delta^\prime - R \frac{\sin \phi}{\cos \delta^\prime}
\end{equation}

For the atmospheric refraction we adopted the classical expression:

\begin{equation}
\label{eq:r}
R = R_0 \;\frac{1 +0.123 h^\prime + 1.255\times10^{-4} h^{\prime 2}}
{1+0.505h^\prime+0.0845 h^{\prime 2}}
\end{equation}

where $R$ is expressed in degrees. The parameter $R_0$ is the
refraction at $h^\prime$=0, which in the above formulation is given
by:

\begin{displaymath}
 R_0 = 0.159 \; \frac{P}{T}
\end{displaymath}

where $P$ is the atmospheric pressure (in mb) and T is the temperature
(in K). For typical values (T=293 K, P=1000 mb) this gives
$R_0\sim$0.5 degrees. Eq.~\ref{eq:r} provides a reasonably good
approximation to the altitude dependence for $h^\prime<$15 degrees
(Sampson, Lozowski \& Machel \cite{sampson05}). However, astronomical
refraction close to the horizon is known to be very variable and one
cannot assume a single constant value (Schaefer \& Liller
\cite{schaefer90}). For instance, Sampson et al. (\cite{sampson03})
report an average value of 0.67 degrees and a standard deviation of
0.17 degrees, with a range of 0.4 to 2.1 degrees. They also show that
there are systematic differences between sunrise and sunset (see also
Schaefer \& Liller \cite{schaefer90}) and that climatic trends play a
role.

To account for this variability, we have adopted the distribution
derived from the Sampson et al. (\cite{sampson03}) data set, which
includes 348 measurements obtained during both sunset and sunrise,
under a wide range of atmospheric conditions (i.e. pressure,
temperature and vertical temperature gradient). The distribution is
plotted in Fig.~A\ref{fig:sampson}.

The rms error on the declination was estimated via Monte Carlo
simulations that included the uncertainties on measured azimuths
($a$), horizon altitudes ($h^\prime$)  and refraction ($R_0$). While
the random realizations of azimuths and altitudes were generated with
a Gaussian distribution, the refractions were drawn from the Sampson
et al. (\cite{sampson03}) distribution. The refraction at altitude
$h^\prime$ was estimated using Equation~\ref{eq:r}.

When predicting the relevant azimuths (e.g. sunrises and sunsets at
solstices and equinoxes), the uncertainties were estimated keeping the
sun declination fixed at its nominal value ($-\varepsilon$, 0,
$+\varepsilon$), while horizon altitudes and refractions were generated
as described in the previous paragraph.

Refraction and horizon altitude corrected values for declinations and
azimuths were derived numerically, using the horizon profile computed
as described in Patat (\cite{patat11}).

\section{\label{sec:appb} Appendix B - Digital Elevation Models}

\begin{figure*}
\centering
\includegraphics[width=17cm]{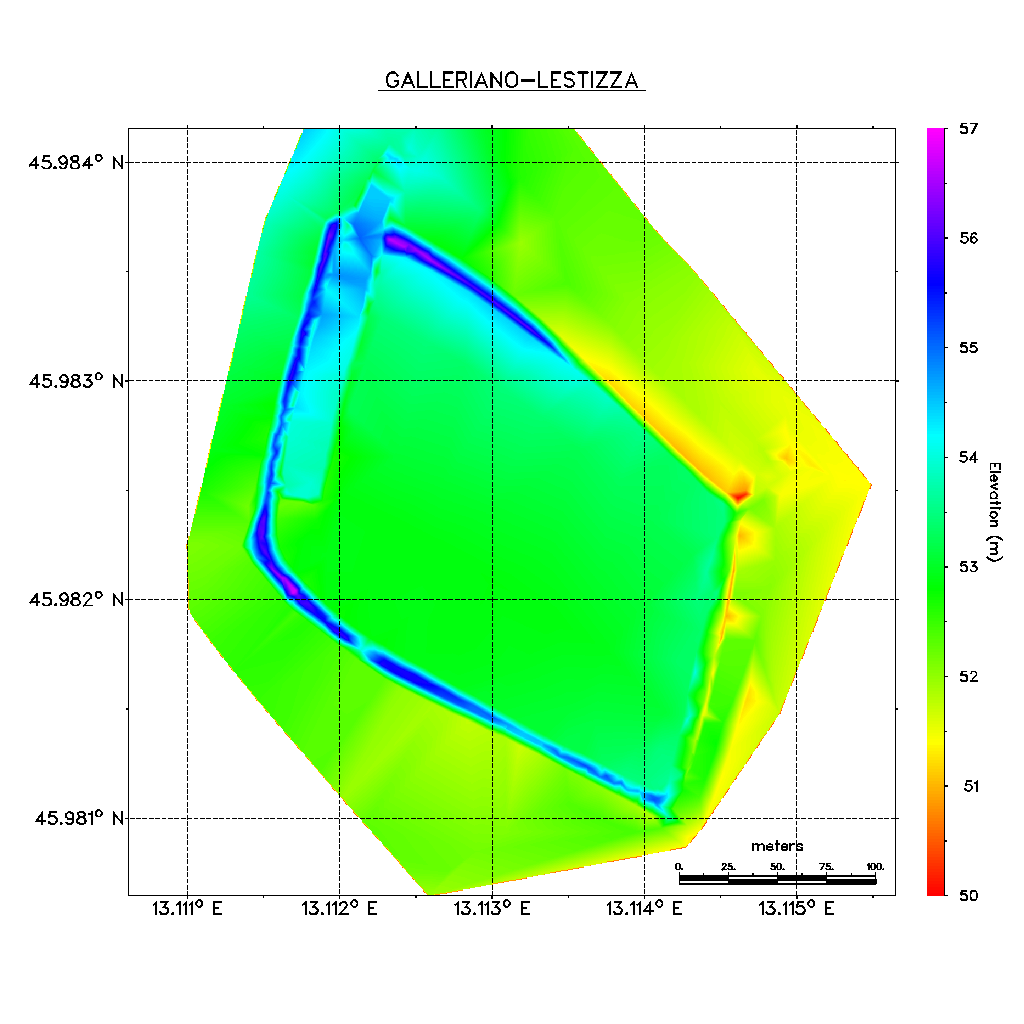}  
\caption{\label{fig:map_gall} Elevation map for the site of Galleriano di Lestizza. Geodetic coordinates are in the WGS84 grid.}
\end{figure*}

\begin{figure*}
\centering
\includegraphics[width=17cm]{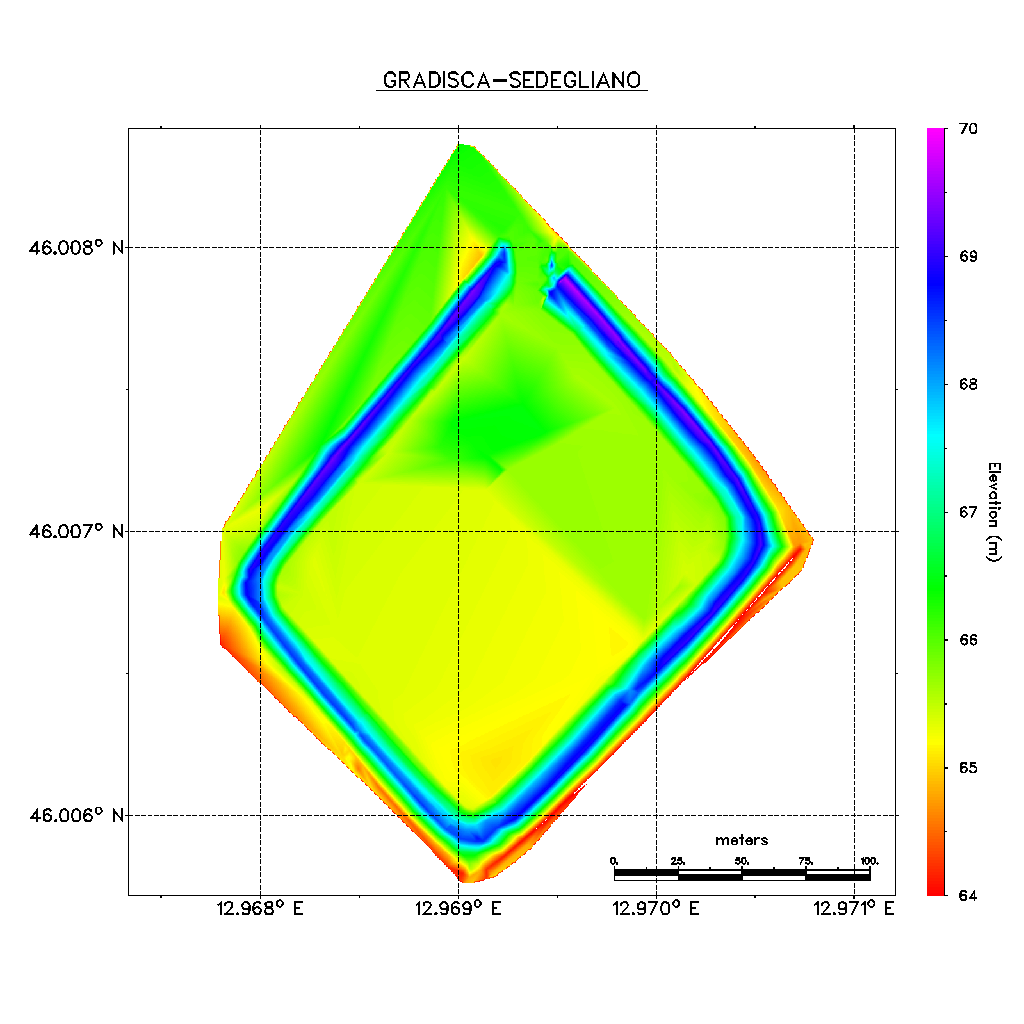}  
\caption{\label{fig:map_grad} Elevation map for the site of Gradisca di Sedegliano. Geodetic coordinates are in the WGS84 grid.}
\end{figure*}

\begin{figure*}
\centering
\includegraphics[width=17cm]{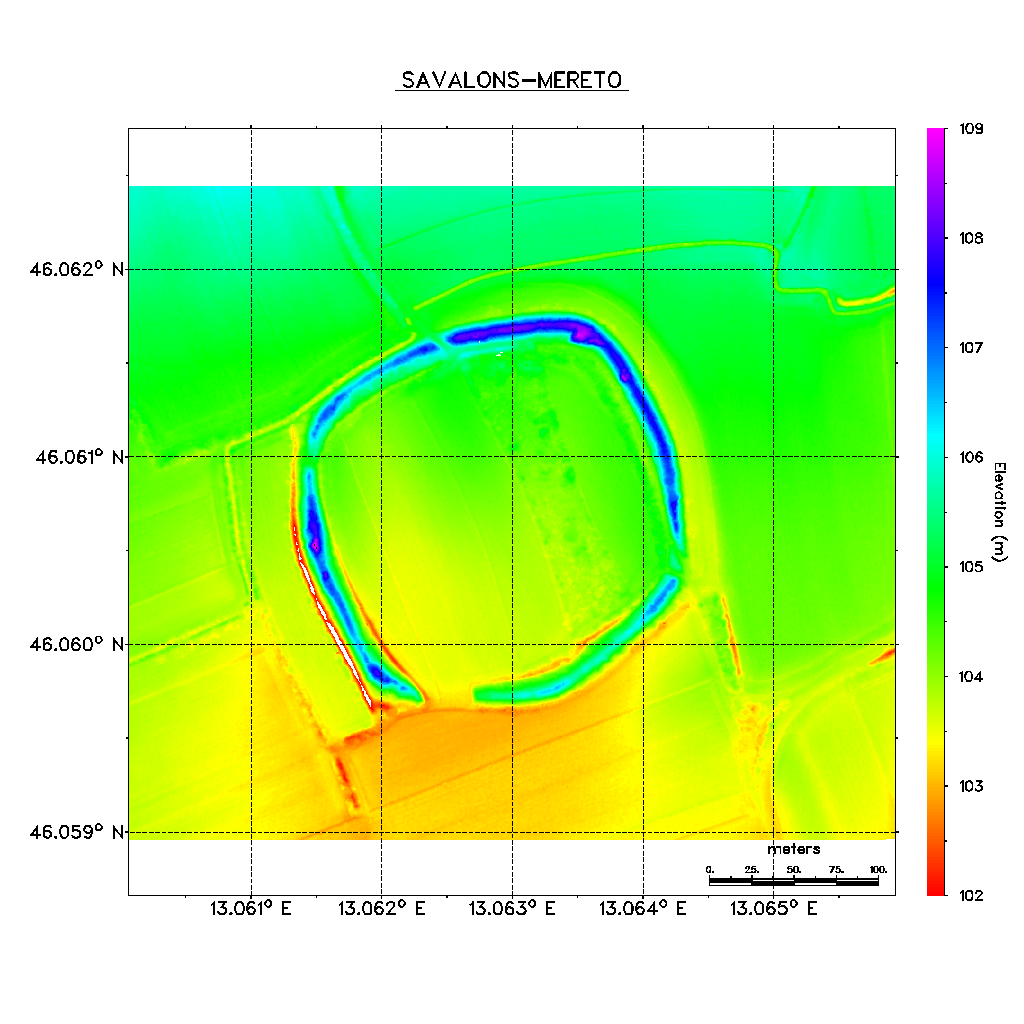}  
\caption{\label{fig:map_sava} Elevation map for the site of Savalons di Mereto. Geodetic coordinates are in the WGS84 grid.}
\end{figure*}

\begin{figure*}
\centering
\includegraphics[width=17cm]{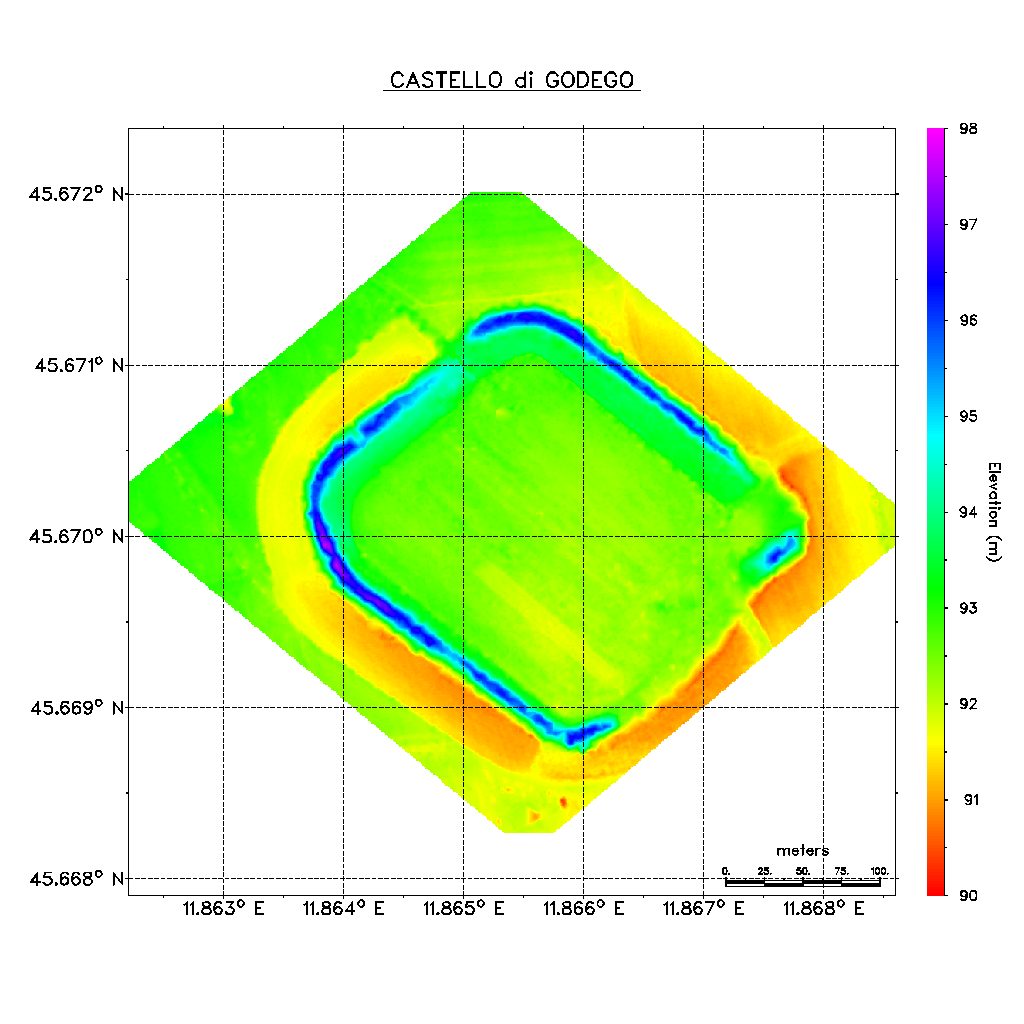}  
\caption{\label{fig:map_cdg} Elevation map for the site of Castello di Godego. Geodetic coordinates are in the WGS84 grid.}
\end{figure*}


\begin{thebibliography}{}
\bibitem[1986]{ar86}Aveni, A.~F. \& Romano, G., Archaeoastronomy 10 
  (Supplement to Journal for the History of Astronomy 17): S24 ({\bf AR86})
\bibitem[1989]{balista} Balista, C., 1989, Quaderni di
  Archeologia del Veneto, V, CEDAM, p. 253
\bibitem[1989]{bc89} Bianchin Citton, E., 1989, Quaderni di
  Archeologia del Veneto, V, CEDAM, p. 216
\bibitem[1991]{bc91} Bianchin Citton, E., 1991, in {\it Colloquio
  internazionale di Archeologia e Astronomia}, G. Romano \&
  G. Traversari Eds., RdA Supplement, no.~9, p.~30
\bibitem[2009]{borgna} Borgna, E. \& C\`assola Guida, P., 2010, in
  {\it A connecting sea: maritime interaction in Adriatic prehistory},
  Staso Forenbaher (Ed.), BAR International Series 2037, 89
\bibitem[2013]{borgna13} Borgna, E., Corazza, S. \& Simeoni, G., 2013,
  in {\it Atti del Forum per la ricerca archeologica nel Friuli-Venezia Giulia},
  Aquileia, January 2011, p. 34
\bibitem[1981]{cg80} C\`assola Guida, P., 1980, in {\it Castelli del
  Friuli}, Vol.~2, Storia ed evoluzione dell'arte delle fortificazioni
  in Friuli, Tito Miotti Ed., p.~13
\bibitem[2006] {cg06}C\`assola Guida, P., 2006, in {\it Tracce
  archeologiche di antiche genti - La protostoria in Friuli}, Circolo
  Culturale Menocchio - Montereale Valcellina. 17-50
\bibitem[2003a]{cgc03a}C\`assola Guida, P. \& Corazza, S., 2003a, {\it Il
  tumulo di Santo Osvaldo}, Universit\`a di Udine (ed.)
\bibitem[2003b]{cgc03b}C\`assola Guida, P. \& Corazza, S., 2003b, Dai
  tumuli ai castellieri 2003b, 74, 650
\bibitem[2003c]{cgc03c}C\`assola Guida, P. \& Corazza, S., 2003c, Dai
  tumuli ai castellieri 2003c, 74, 654
\bibitem[2004a]{cgc04a}C\`assola Guida, P. \& Corazza, S., 2004a, Dai
  tumuli ai castellieri 2004, 75, 547
\bibitem[2004b]{cgc04b}C\`assola Guida, P. \& Corazza, S., 2004b, Dai
  tumuli ai castellieri 2004, 75, 525
\bibitem[2005]{cgc05}C\`assola Guida, P. \& Corazza, S., 2005, in {\it
  Carlo Marchesetti e i castellieri 1903-2003}, G. Bandelli and
  E. Montagnari Kokelj (eds.), 221
\bibitem[2009]{cgc09} C\`assola Guida, P. \& Corazza, S., 2009, in
  {\it Notiziario della Soprintendenza per i Beni Archeologici del
    Friuli-V.G.}, 2/2007, 144
\bibitem[2000]{susi00} Corazza, S., 2000, Aquileia Nostra, 71, 645
\bibitem[1976]{caporiacco} di Caporiacco, G., 1976, {\it Udine e il
  suo territorio. Dalla preistoria alla latinit\`a}, Arti Grafiche
  Friulane, Udine, 11
\bibitem[2007]{farr} Farr, T.~G., et al.: 2007, Rev. Geophys. 45,
  RG2004
\bibitem[2000]{lupato00}Lupato, G., 2000, Astronomia UAI,
          N. 1, 3
\bibitem[1988]{menis} Menis, G.~C., 1988, {\it History of Friuli},
  Grafiche Editoriali Artistiche Pordenonesi, Pordenone, 16
\bibitem[2011]{patat11} Patat, F., 2011, AN, 332, 743
\bibitem[2009]{pimenta} Pimenta, F., Tirapicos, L. \& Smith, A., 2009,
  Archaeoastronomy, 22, 1
\bibitem[2002]{prenc02} Prenc, F., 2002, Antichit\`a Altoadriatiche, 52
\bibitem[1943]{quarina}Quarina, L., 1943, Ce fastu?, 19. 54-86
\bibitem[1981]{romano81} Romano, G., 1981, Mem. S.A.It, 52, 627
\bibitem[1992]{romano}Romano, G., 1992, {\it Archeoastronomia Italiana},
	Cooperativa Libraria Editrice Universit\`a di Padova, 222
\bibitem[2003]{sampson03} Sampson, R.~D., Lozowski, E.~P., Peterson,
  A.~E. \& Hube, D.~P., 2003, PASP, 115, 1261
\bibitem[2005]{sampson05} Sampson, R.~D., Lozowski, E.~P. \& Machel,
  H.~G., 2005, Applied Optics. 44, 5652
\bibitem[1990]{schaefer90} Schaefer, B.~E. \& Liller, W., 1990, PASP,
  102, 796
\bibitem[2011]{simeoni} Simeoni G. \& Corazza, S., 2011, {\it Di terra
  e di ghiaia - La protostoria del Medio Friuli tra Europa e
  Adriatico}, La Grame (Ed.)
\end{thebibliography}
\end{document}